\documentclass[12pt,a4paper]{article}
\usepackage[utf8]{inputenc}
\usepackage{a4wide}
\usepackage{graphicx}
\newcommand{\be}{\begin{equation}}
\newcommand{\ee}{\end{equation}}
\newcommand{\ba}{\begin{eqnarray}}
\newcommand{\ea}{\end{eqnarray}}
\newcommand{\Tr}{\mathrm{Tr}}
\newcommand{\Str}{\mathrm{Str}}

\newcommand{\nn}{\nonumber\\}
\newcommand{\lgl}{\langle}
\newcommand{\rgl}{\rangle}
\begin{document}

\begin{titlepage}
\begin{flushright}
LU TP 15-29\\
August 2015
\end{flushright}
\vfill
\begin{center}
{\Large\bf Finite Volume for Three-Flavour\\[4mm]
Partially Quenched Chiral Perturbation Theory\\[6mm]
through NNLO in the Meson Sector}

\vfill

{\bf Johan Bijnens and Thomas Rössler}\\[0.3cm]
{Department of Astronomy and Theoretical Physics, Lund University,\\
S\"olvegatan 14A, SE 223-62 Lund, Sweden}
\end{center}
\vfill
\begin{abstract}
We present a calculation of the finite volume corrections to meson masses
and decay constants in three flavour Partially Quenched Chiral Perturbation
Theory (PQChPT) through two-loop order in the chiral expansion for the
flavour-charged (or off-diagonal) pseudoscalar mesons. The analytical
results are obtained
for three sea quark flavours with one, two or three different masses.
We reproduce the known infinite volume results and the finite volume results
in the unquenched case. The calculation has been
performed using the supersymmetric formulation of
PQChPT as well as with a quark-flow technique.

Partial analytical results can be found in the appendices.
Some examples of cases relevant to lattice QCD are studied numerically.
Numerical programs for all results are
available as part of the \textsc{CHIRON} package.
\end{abstract}
\vfill
\vfill
\end{titlepage}


\section{Introduction}
\label{sec:Introduction}

Quantum chromodynamics (QCD) is nowadays accepted to be the theory describing
the strong force. 
The smallness of the coupling constant at high energies makes it possible to
test and confirm the theory in highly energetic scattering. It also provides
 -- at least in principle -- a way to obtain various low-energy hadronic
observables, such as masses and decay constants, but it has hitherto been
impossible to derive such quantities of interest in terms of analytical
expressions by means of ab initio calculations. A numerical approach that can
circumvent the problem is lattice QCD.
A review of the applications to flavour and low-energy hadron physics
is \cite{Aoki:2013ldr}. 
To calculate observables, one uses a numerical evaluation of the QCD path
integral in a Monte Carlo approach.
A number of restrictions follow from the nature of the calculation.
Since it is carried out on a space-time lattice in a finite volume, it is of
high interest to have the effect of the finite volume
under good control. Furthermore, although lattice computations in meson
physics are now feasible when using physical parameters for the light quark
masses, a lot of calculations still use unphysically high masses for the quarks.
It is also useful to vary quark masses to study a number of phenomena.
A common solution to study quark mass dependence with lower computational
needs is given by partial quenching. In partially quenched QCD (PQQCD), one
associates different masses (usually larger ones) to the sea quarks and
the valence quarks. Valence quarks are those connected to the external operators
while sea quarks are those in the fermion determinant or equivalently in
closed loops. Sea quarks are only connected to external states via gluons. 

The preferable way to correct for unphysical quark masses
is by means of Chiral Perturbation Theory (ChPT)
\cite{Weinberg:1978kz,Gasser:1983yg,Gasser:1984gg}. Finite volume effects
for ChPT have been introduced in
\cite{Gasser:1986vb,Gasser:1987ah,Gasser:1987zq}. The corresponding effective
theory for PQQCD is given by Partially Quenched Chiral Perturbation Theory
(PQChPT) \cite{Bernard:1993sv}. The arguments underlying this
are elaborated in \cite{Bernard:2013kwa}.

The proper matching of calculations in PQChPT to results from Partially
Quenched Lattice QCD allows a whole new landscape of possibilities, such as
improved validation and extrapolation of lattice results, or a more accurate
determination of the chiral low-energy constants (LECs), see
e.g. \cite{Sharpe:2000bc}. It should be stressed that, as opposed to fully
quenched calculations, partially quenched calculations are connected to
their corresponding unquenched scenarios by a \emph{continuous} change in
variables, making it possible to immediately extract physical results from
otherwise unphysical simulations.

In this paper, we address the finite volume corrections through two-loop
order in the PQChPT framework, specifically for the flavour-charged or off-diagonal
mesons. The infinite volume (IV) results in PQChPT to this order are known
for three \cite{Bijnens:2004hk,Bijnens:2005ae,Bijnens:2006jv} and
two \cite{Bijnens:2005pa} sea quark flavours. 
The finite volume (FV) corrections in (unquenched) ChPT at two-loop
order have been addressed
in our earlier study \cite{Bijnens:2014dea}. The needed integrals have been
worked out in \cite{Bijnens:2013doa}.
Our expressions are valid in the frame with vanishing spatial momentum,
$\vec p=0$, often
called the center-of-mass frame. In the so-called moving frames
or with twisted boundary conditions there will be additional terms.
We have chosen to present our result in terms of lowest order
masses given the ambiguity in expressing the
results in terms of the large number of possible different physical masses.

Earlier work on finite volume corrections at NNLO are besides our own
work \cite{Bijnens:2014dea}, the pion mass in two-flavour
ChPT \cite{Colangelo:2006mp} and the vacuum expectation value
in three flavour ChPT \cite{Bijnens:2006ve}. Extensions of the latter
work to partially quenched are in \cite{Damgaard:2008zs}.
We did not find published results for the finite volume corrections
at one-loop order in the partially quenched case. They are however implicit
in the expressions given for the staggered partially quenched case
in \cite{Aubin:2003mg,Aubin:2003uc}.

We give a short list of references for ChPT and discuss some small points in
Sect.~\ref{sec:PQChPT}. The definitions of the integrals we use and how they
relate to the results in \cite{Bijnens:2013doa} is given in
Sect.~\ref{sec:finitevolumeintegrals}. The next section describes our
major result which is the full finite volume correction to the pion mass
and decay
constant to two-loop order in ChPT. Sect.~\ref{sec:analytical} contains the
results for the three-flavour case for pion, kaon and eta for both the
mass and decay constant. The large two-loop order formulas are
collected for one case in the appendices and all of them can be downloaded
from \cite{chptweb}. A numerical discussion of our results
is in Sect.~\ref{sec:numerical}.

\section{Partially Quenched Chiral Perturbation Theory}
\label{sec:PQChPT}

This section is very similar to the description of PQChPT given in
\cite{Bijnens:2006jv} since our work is the extension to finite volume
of that paper.

An introduction to ChPT can be found in \cite{Scherer} and in the
two-loop review \cite{Bijnens:2006zp}. The lowest order and $p^4$-Lagrangian
can be found in \cite{Gasser:1984gg}.
The order $p^6$ Lagrangian is given
in \cite{Bijnens:1999sh}. We use the standard renormalization scheme in ChPT.
An extensive discussion of the renormalization scheme can
be found in \cite{Bijnens:1997vq} and \cite{Bijnens:1999hw}.
Important for our work is that the LECs do not depend on the
volume \cite{Gasser:1987zq}.
An introduction with applications to lattice QCD is \cite{Golterman:2009kw}.
References to more introductory literature can be found on \cite{chptweb}.

The expansion in ChPT is in momenta $p$ and quark-masses. We count
the latter as two powers of $p$. This counting is referred to as
$p$-counting.
We prefer to designate orders by the $p$-counting order at which the diagram
appears. Thus we refer to lowest order (LO) as order $p^2$,
next-to-leading order (NLO) as order $p^4$ or one-loop order
and next-to-next-to-leading order (NNLO) as order $p^6$ or two-loop order and
include in the terminology one- or two-loop order also the diagrams with
fewer loops but the same order in $p$-counting.

\subsection{The Lagrangian}

Three massless quark flavours QCD has a chiral symmetry
\begin{equation}
G=SU(n_f)_L\times SU(n_f)_R\,.  
\label{chsym}
\end{equation}
which is spontaneously broken to the diagonal subgroup
$SU(3)_V$. The Goldstone bosons following from this spontaneous breakdown
are described by the meson octet matrix
\begin{eqnarray}
\label{Phi}
\phi(x)&=
\left(\begin{array}{ccc}
\frac{1}{\sqrt{2}}\pi^0+\frac{1}{\sqrt{6}}\eta &\pi^+&K^+\\
\pi^-&-\frac{1}{\sqrt{2}}\pi^0+\frac{1}{\sqrt{6}}\eta&K^0\\
K^- &\bar{K}^0&-\frac{1}{\sqrt{3}}\eta
\end{array}\right),
\end{eqnarray}
The flavour-singlet component has been integrated out
since it is heavy due to the $U(1)_A$ anomaly.
The spontaneous symmetry breaking is the basis of ChPT.

In partially quenched QCD one distinguishes between valence and sea quarks.
Valence quarks are connected to the external states (or operators) while
the sea-quarks are those contributing in closed loops only connected
via gluons to external states. These can be given different masses
in lattice QCD calculations. The ChPT for this
partial quenching can be done by studying the quark flow generalizing the
quenched case studied in \cite{Sharpe:1992ft}. One can then treat
the the sea and valence lines differently.
Alternatively, one can make use of the supersymmetric formulation of PQChPT
\cite{Bernard:1993sv}. In the latter, three corresponding sets of quarks are
introduced instead of only two: In addition to the valence and sea sector,
a set of so-called ghost quarks is added. These are ``bosonic'' in the sense
that they are treated as commuting variables. With their masses fixed to the
same numerical values as present in the valence sector, they will cancel
exactly the contribution coming from closed valence quark loops.
Most of the remainder of this section will be concerned with the supersymmetric
formulation. The changes needed to use a quark flow technique are discussed
at the end.

The chiral symmetry group is formally extended to the graded\footnote{The
precise structure of the symmetry group is somewhat different, but the one
given here is sufficient for both the present discussion as well as for
practical calculations. The ``approximate'' symmetry group reproduces the
right Ward identities \cite{Sharpe:2000bc,Sharpe:2001fh}.}
\begin{equation}
G = SU(n_\mathrm{val}+n_\mathrm{sea} | n_\mathrm{val})_L
\times SU(n_\mathrm{val}+n_\mathrm{sea} | n_\mathrm{val})_R\,,
\end{equation}
for the case of $n_\mathrm{val}$ valence and $n_\mathrm{sea}$ quarks.
The chiral group $G$ is spontaneously broken to the diagonal subgroup
$SU(n_\mathrm{val}+n_\mathrm{sea} | n_\mathrm{val})_V$.
We will work in the flavour basis rather than in the meson basis.
We will thus use fields $\phi_{ab}$ corresponding to the flavour content
of $q_a\bar q_b$. The mixing of the neutral eigenstates and the
integrating out of the singlet degree of freedom is taken
care of by using a more complicated propagator\footnote{This is
described in detail in \cite{Sharpe:2001fh}. It is possible to use the same
method also in standard ChPT.}.

The corresponding Goldstone degrees of freedom are put in a matrix
with the generic structure
\begin{equation}
\label{SUSY_FieldMatrix}
\Phi =
\left(\begin{array}{ccc}
\Big[\;\;q_V\bar q_V\;\;\Big] & 
\Big[\;\;q_V\bar q_S\;\;\Big] &
\Big[\;\;q_V\bar q_B\;\;\Big] \\ \\ 
\Big[\;\;q_S\bar q_V\;\;\Big] &
\Big[\;\;q_S\bar q_S\;\;\Big] & 
\Big[\;\;q_S\bar q_B\;\;\Big]\\ \\
\Big[\;\;q_B\bar q_V\;\;\Big] & 
\Big[\;\;q_B\bar q_S\;\;\Big] &
\Big[\;\;q_B\bar q_B\;\;\Big]
\end{array}\right)\,.
\label{pqmatr}
\end{equation}
$V$ denotes valence, $S$ denotes sea and $B$ denotes the bosonic ghost quarks.
Note that the meson fields containing one single ghost quark only will
themselves obey fermionic, i. e. anticommuting, statistics.

The structure of the Lagrangian is similar to standard ChPT for a generic
number of flavours.
The lowest order Lagrangian is
\begin{equation}
\label{L2}
{\cal L}_2 = \frac{F_0^2}{4}\lgl u_\mu u^\mu + \chi_+\rgl\,. 
\end{equation}
At one-loop, it is given by
\begin{eqnarray}
\label{L4}
{\cal L}_4 &=&
 {\hat L}_0\,\lgl u^\mu u^\nu u_\mu u_\nu \rgl 
+{\hat L}_1\,\lgl  u^\mu u_\mu \rgl^2 
+{\hat L}_2\,\lgl u^\mu u^\nu \rgl \lgl u_\mu u_\nu \rgl
+{\hat L}_3\,\lgl (u^\mu u_\mu)^2 \rgl
\nn &&
+ {\hat L}_4\,\lgl u^\mu u_\mu \rgl \lgl \chi_+\rgl 
+ {\hat L}_5\,\lgl u^\mu u_\mu \chi_+ \rgl 
 +{\hat L}_6\,\lgl \chi_+ \rgl^2 
+ {\hat L}_7\,\lgl \chi_- \rgl^2
+ \frac{{\hat L}_8}{2}\,\lgl \chi_+^2 + \chi_-^2 \rgl 
+ \ldots\,.
\end{eqnarray}
We show only the terms relevant for our work.

The generalized Goldstone manifold is parametrized by
\begin{equation}
u\equiv\exp\left(i\Phi/(\sqrt{2}\hat F)\right)
\end{equation}
similar to the exponential representation in standard ChPT. It is a
$9\times9$ matrix with fermionic parts.
We have
furthermore introduced
\begin{eqnarray}
u_\mu &=& i\left\{
u^\dagger(\partial_\mu-i r_\mu)\,u -
u\,(\partial_\mu-i l_\mu)\,u^\dagger\right\},
\nonumber \\
\chi_\pm &=& u^\dagger\chi\,u^\dagger\pm u\,\chi^\dagger\,u\,.
\label{uquant}
\end{eqnarray}
The matrix $\chi$ is for this work restricted to
\begin{equation}
\chi = 2 B_0\,\mathrm{diag}(m_1,\ldots,m_9)\,
\end{equation}
with $m_i$ the quark mass of quark $i$ and $B_0$ a LEC. 
We have here $m_1=m_7, m_2=m_8, m_3=m_9$ as the valence
masses and $m_4,m_5,m_6$ as the sea quark masses.
Ordinary traces have been replaced by supertraces, denoted by
$\lgl~\rgl$, defined in terms of the ordinary ones by
\begin{equation}
\Str \left(\begin{array}{cc} A & B \\ C & D \end{array}\right)
=\Tr\,A - \Tr\,D\,.
\end{equation}
$B$ and $C$ denote the fermionic blocks in the matrix.
The supersinglet $\Phi_0$, generalizing the $\eta'$, is integrated out
to account for the axial anomaly as in standard ChPT,
implying the additional condition
\begin{equation}
\label{tracezero}
\lgl \Phi \rgl = \Str\,(\Phi) = 0\,.
\end{equation}
However, as mentioned above, we will work in the flavour basis enforcing
the constraint (\ref{tracezero}) via the propagator.

A calculation in PQChPT has to be performed using a larger set of operators
since no further reduction by means of Cayley-Hamilton relations can be
performed. The three-flavour PQChPT Lagrangian (equation (\ref{L4})) thus
has 11 LECs for PQChPT.

The LECs for standard three flavour ChPT
are related to those of three flavour PQChPT via
\begin{equation}
L_1^r = \hat L_1^{r} + \hat L_0^{r}/2,\qquad
L_2^r = \hat L_2^{r} + \hat L_0^{r},\qquad
L_3^r = \hat L_3^{r} - 2\,\hat L_0^{r},
\end{equation}
and $L_i^r=\hat L_i^r$ for the others.
Note that a numerical value for $\hat L_0$ cannot be obtained by experiment,
but can be determined only via PQQCD lattice simulations or modelling.

An additional comment is that the divergences for PQChPT are directly
related to those for $n_{sea}$-flavour ChPT \cite{Bijnens:1999hw}
when all traces are replaced by supertraces. This can be argued using
the formal equivalence of the equations of motion used or via the
replica trick \cite{Damgaard:2000gh}.

\subsection{The propagator and notation for masses and residues}
\label{sec:Notation}

The variant of PQChPT, considered in this paper, comes with three valence
quarks, with masses $m_1,m_2,m_3$ and three sea quarks with
masses $m_4,m_5,m_6$.
The additional ghost quarks emerging only in the supersymmetric formulation
have masses $m_7,m_8,m_9$.
They do not appear explicitly since they are fixed to the ones in the valence
sector, i.e. $m_7 = m_1$, $m_8 = m_2$, $m_9 = m_3$.

We use the numbers $d_{val}$ and $d_{sea}$ to denote the number of
non-degenerate quark masses in each sector. In the case of two non-degenerate
mass scales for one sector, it is the two masses with the lowest indices
that we set degenerate, which will in turn both be represented by the mass
scale with the lowest index, e.g. in the case $d_{sea}=2$ we have
$m_4 = m_5 \neq m_6$ and expressions will be explicitly dependent on
$m_4$ and $m_6$ only.

In fact we will always absorb a factor $2B_0$ in the notation and we use
\begin{equation}
\chi_i\equiv 2 B_0 m_i\,,\qquad
\chi_{ij}\equiv \frac{1}{2}\left(\chi_i+\chi_j\right)\,.
\end{equation}
The lowest order masses for off-diagonal mesons 
with flavour content $q_i\bar q_j$ are given by $\chi_{ij}$
and we will use $\chi_i$ rather then $\chi_{ii}$
for equal masses. Dealing with masses for the diagonal valence
mesons in PQChPT is not trivial.
This is discussed in detail in \cite{Sharpe:2001fh} and extended to NNLO in
\cite{Bijnens:2006ca}.
The diagonal sea quark sector has two masses associated with it,
corresponding to
the neutral pion and eta masses.
These we denote by $\chi_\pi$ and $\chi_\eta$.
They are defined as the solutions to the equations
\begin{eqnarray}
\label{neutral_sea_masses}
\chi_\pi+\chi_\eta &=& \frac{2}{3}\left(\chi_4+\chi_5+\chi_6\right),
\nonumber\\
\chi_\pi \chi_\eta &=& \frac{1}{3}
\left(\chi_4\chi_5+\chi_5\chi_6+\chi_4\chi_6\right).
\end{eqnarray}
They are non-polynomial in the sea masses $\chi_j$ for three
non-degenerate quark masses, i.e. $d_{sea}=3$.
For $d_{sea}=2$ one has instead $\chi_\pi=\chi_4$ and 
$\chi_\eta=(1/3)(\chi_4+2\chi_6)$.

The flavour-charged propagator, connecting $\phi_{ij}$
with $\phi_{ji}$, is given by
\cite{Bernard:1993sv,Sharpe:2000bc,Sharpe:2001fh}
\begin{eqnarray}
-i\,G_{ij}^c (k) &=& 
\frac{\epsilon_j}{k^2 - \chi_{ij} + i\varepsilon}\quad (i \neq j)\,.
\label{propc}
\end{eqnarray}   
with $\chi_{ij} \equiv (\chi_i + 
\chi_j) / 2$, the lowest order meson mass, and the 
signature $\epsilon_j$ is defined as $+1$ for the flavor indices 
of the $n_\mathrm{val} + n_\mathrm{sea}$ fermionic quarks, and as $-1$ 
for the flavor indices of the $n_\mathrm{val}$ bosonic ghost quarks. In 
the present calculation, with the number of valence and sea quarks as 
given above, $\epsilon_j$ thus takes the values
\begin{equation}
\epsilon_j=\left\{
  \begin{array}{cl}
    +1 &  \mathrm{for} \;\;\;j=1,\dots,6\\
    -1 & \mathrm{for}\;\;\;j=7,8,9\,.
  \end{array}
\right.
\end{equation}

The flavour-neutral propagator,
connecting a flavour field $\phi_{ii}$ to $\phi_{jj}$,
on the other hand suffers from additional contributions emerging from
the elimination of the $\Phi_0$
and the partial quenching \cite{Bernard:1993sv,Sharpe:2000bc,Sharpe:2001fh}.
We write it as
\begin{eqnarray}
G_{ij}^n (k) &=& G_{ij}^c (k)\,\delta_{ij} 
- G_{ij}^q (k) / n_\mathrm{sea}.
\label{propn}
\end{eqnarray}
The additional terms are either
\begin{eqnarray}
-i\,G_{ij}^q (k) &=& \frac{R^{i}_{j\pi\eta}}{k^2 - \chi_i + i\varepsilon}
+ \frac{R^{j}_{i\pi\eta}}{k^2 - \chi_j + i\varepsilon} 
\nonumber \\ 
&+& \frac{R_{\eta ij}^\pi} {k^2 - \chi_\pi + i\varepsilon} 
+ \frac{R_{\pi ij}^\eta}{k^2 - \chi_\eta + i\varepsilon},
\label{npropij} 
\end{eqnarray}
for the case with $i\ne j$ and $\chi_i\ne\chi_j$, or
\begin{eqnarray}
-i\,G_{ij}^q (k) &=& \frac{R^d_i}{(k^2 - \chi_i + i\varepsilon)^2} 
+ \frac{R^c_i}{k^2 - \chi_i + i\varepsilon} 
\nonumber \\
&+& \frac{R_{\eta ii}^\pi} {k^2 - \chi_\pi + i\varepsilon} 
+ \frac{R_{\pi ii}^\eta}{k^2 - \chi_\eta + i\varepsilon},
\label{npropij2} 
\end{eqnarray}
for the case with $\chi_i = \chi_j$ which clearly includes $i=j$. 
In the second case, the sum of single poles is supplemented with an
unphysical double pole. Since double poles emerge due to the partial quenching
in the valence sector, they disappear by taking the appropriate unquenched
limit.

Using the ratios of products of differences of masses
\begin{eqnarray}
R^z_{ab} &=& \chi_a - \chi_b, \nonumber \\
R^z_{abc} &=& \frac{\chi_a - \chi_b}{\chi_a - \chi_c}, \nonumber \\ 
R^z_{abcd} &=& \frac{(\chi_a - \chi_b)(\chi_a - \chi_c)}
{\chi_a - \chi_d}, \nonumber \\
R^z_{abcdefg} &=& \frac{(\chi_a - \chi_b)(\chi_a - \chi_c)(\chi_a - \chi_d)}
{(\chi_a - \chi_e)(\chi_a - \chi_f)(\chi_a - \chi_g)},
\label{RSfunc}
\end{eqnarray}
the residues $R$ of the neutral meson propagator
in equations (\ref{npropij}) and (\ref{npropij2}) are (for $d_{sea}=3$)
\begin{equation}
R_{jkl}^{i} = R^z_{i456jkl}, \quad
R_{i}^{d} = R^z_{i456\pi\eta}, \quad
R_{i}^{c} = R^i_{4\pi\eta} + R^i_{5\pi\eta} + R^i_{6\pi\eta}
          - R^i_{\pi\eta\eta} - R^i_{\pi\pi\eta}.
\end{equation}
Note that many of these quantities vanish when $i$ takes the value of
a sea quark index. The sea-quark propagators thus do not contribute any
double poles as expected since these originate from the quenching in the
valence sector.

For $d_{sea}=2$ or $\chi_\pi=\chi_5=\chi_4$.
The needed residues simplify to
\begin{equation}
R_{jk}^{i} = R^z_{i46jk},\quad
R_{i}^{d} = R^z_{i46\eta},\quad
R_{i}^{c} = R^i_{4\eta} + R^i_{6\eta} - R^i_{\eta\eta}.
\end{equation}
The corresponding propagator can be obtained by removing all pion indices
as well as the pion mass pole from equations (\ref{npropij}) and
(\ref{npropij2}).

The physically less interesting case $d_{sea}=1$ immediately yields $\chi_\pi=\chi_\eta=\chi_6=\chi_5=\chi_4$. All residues from the sea quark sector are
reduced to numbers, only 
\begin{equation}
R_{j}^{i} = R^z_{i4j}, \quad \\
R_{i}^{d} = R^z_{i4}\,,
\end{equation}
appear.

\subsection{The quark flow case}

We have performed the calculation using the supersymmetric method described
above but also with the quark flow method \cite{Sharpe:1992ft}.
We use the same Lagrangians as in (\ref{L2}) and (\ref{L4}) but with
normal traces everywhere.
The matrix $\Phi$ is now written in terms of generic fields
$\phi_{ij}$ and all indices are kept symbolic implying summations.

Connecting propagators of a field $\phi_{ij}$ to $\phi_{kl}$
should be done by using
\begin{equation}
G_{ijkl}(k) = G^c_{ij}(k)\delta_{il}\delta_{jk}-\delta_{ij}\delta_{kl}
G^q_{ik}(k)/n_\mathrm{sea}\,.
\end{equation}
The propagators $G^c_{ij}(k),G^q_{ij}(k)$ remain the same but we can now
disregard the factors $\epsilon_j$ since with this method there are no
bosonic ghost quarks.

After constructing the Feynman diagrams using the above,
the quark flow is visible following the symbolic flavour indices.
Next, one replaces the index lines
that connect to external fields or operators by their appropriate
valence value. The remaining index lines are now sea indices and
are summed over with the sea quark indices.

The results obtained with the quark flow method agreed in all cases
with those of the supersymmetric method.

\section{The finite volume integrals}
\label{sec:finitevolumeintegrals}

The loop integrals at finite volume at one-loop are well known.
There is a sum over discrete
momenta in every direction with a finite size rather than a continuous integral.
The Poisson summation formula allows to identify the
infinite volume part and the finite volume corrections.
The remainder can be done with two different methods. For one-loop tadpole
integrals the first method was introduced by 
\cite{Gasser:1986vb,Gasser:1987ah,Gasser:1987zq} and 
a sum over Bessel functions, that for large $ML$ converges fast, remains to be
done.
With the other method one remains instead with
an integral over a Jacobi theta function, this method can be used for small
and medium $ML$ as well. It can be found in \cite{Becirevic:2003wk}.
 The extensions to other one-loop
integrals is done in both cases by combining propagators with Feynman
parameters. The first method was extended to the equal mass two-loop
sunset integral \cite{Colangelo:2006mp} and later
to the more general mass case in \cite{Bijnens:2013doa}. 
The latter extended the Jacobi theta function method as well to the sunset
case. Details and further references can be found in \cite{Bijnens:2013doa}.
In this paper we use Minkowski notation for the integrals.

For the one-loop integrals needed here, we use a notation that does a first
classification according to the sum of the powers of the propagators
with different masses, $m_1, m_2, ..., m_{max}$.
We label the integrals $A,B,C,D$ for a total power of
propagators of $n=1,2,3,4$ respectively, since total powers of up to 4
can appear in the calculation as follows from the discussion of double poles
in Sec.~\ref{sec:Notation}.  The different mass scales are given
as consecutive arguments of the integral. Alternatively, if only one mass
scale in total is present, we omit its repetition as a shorthand notation.
For the present calculation at most two different scales
can appear.

Both scalar and tensor integrals will occur, e. g. in the simplest case
of one single propagator raised to single power
\be
\left\{A(m^2),A_{\mu\nu}(m^2)\right\}
 = \frac{1}{i}\int_V\frac{d^d r}{(2\pi)^d}
\frac{\left\{1,r_\mu r_\nu\right\}}{(r^2-m^2)}\,.
\ee
We used the subscript $V$ to indicate it is a finite volume sum and integral.

More Lorentz structures are possible than in the infinite volume case.
We define the tensor $t_{\mu\nu}$ as the spatial part of the Minkowski metric
$g_{\mu\nu}$, to express these. For the center-of-mass (cms) case this is
sufficient. The needed functions for the above example are
\be
A_{\mu\nu}(m^2) = g_{\mu\nu}A_{22}(m^2)+t_{\mu\nu}A_{23}(m^2)\,.
\ee
We then use Passarino-Veltman identities in order to further simplify the
result. In infinite volume the relation obtained by considering
$g^{\mu\nu}A_{\mu\nu}(m^2)$ can be used to remove $A_{22}$.
In finite volume, we again remove the $A_{22}$-type integrals from the
extended relation
\be
dA_{22}(m^2)+3A_{23}(m^2) = m^2 A(m^2)\,.
\ee

Each integral is split into an infinite volume contribution and a finite
volume correction by means of the Poisson summation formula, while
simultaneously being expanded in $\epsilon$ up to the necessary order.
\be
A(m^2) = \lambda_0 \frac{m^2}{16\pi^2}+\overline A(m^2)+A^V(m^2)
+\epsilon\left(A^\epsilon(m^2)+A^{V\epsilon}(m^2)\right)+\cdots\,.
\ee
Here, $\lambda_0=\frac{1}{\epsilon}+\log(4\pi)+1-\gamma$.
The same split is done for all one-loop integrals.
The expressions can be obtained by using the relations
\ba
B(m^2)&=& \frac{\partial}{\partial m^2} A(m^2),
\nonumber\\
C(m^2) &=&\frac{1}{2} \frac{\partial}{\partial m^2} B(m^2),
\nonumber\\
D(m^2) &=&\frac{1}{3} \frac{\partial}{\partial m^2} C(m^2),
\nonumber\\
B(m_1^2,m_2^2) &=&\frac{A(m_1^2)-A(m_2^2)}{m_1^2-m_2^2}.
\ea

The sunset integrals, defined as
\ba
\label{defH}
\lefteqn{\left\{H,H_\mu,H_\mu^s,H_{\mu\nu},H_{\mu\nu}^{rs},H_{\mu\nu}^{ss}\right\}
(n,m_1^2,m_2^2,m_3^2,p) =}
&&\nonumber\\&&
\frac{1}{i^2}\int_V\frac{d^d r}{(2\pi)^d}\frac{d^d s}{(2\pi)^d}
\frac{\left\{1,r_\mu,s_\mu,r_\mu r_\nu, r_\mu s_\nu, s_\mu s_\nu\right\}}
{\left(r^2-m_1^2\right)^{n_1}\left(s^2-m_2^2\right)^{n_2}\left((r+s-p)^2-m_3^2\right)^{n_3}}\,,
\ea
now come with eight different pole configurations. We label these by the
index $n$ according to Tab.~\ref{tab:powers} analoguous to
the infinite volume definitions of
\cite{Bijnens:2004hk,Bijnens:2005ae,Bijnens:2006jv,Bijnens:2005pa}.
\begin{table}
\begin{center}
\begin{tabular}{c|c c c}
 & \quad $ n_1 $ \quad & \quad $ n_2 $ \quad & \quad $ n_3 $ \quad \\ & 
\vspace{-.2cm} \\
\hline\hline
& \vspace{-.2cm} \\
$n=1$ \quad & \quad 1 \quad & \quad 1 & \quad 1\quad\\
\vspace{-.2cm} &&& \\ \hline \vspace{-.2cm} &&& \\
$n=2$ \quad & \quad 2 \quad & \quad 1 & \quad 1\quad\\
$n=3$ \quad & \quad 1 \quad & \quad 2 & \quad 1\quad\\ 
($n=4$) \quad & \quad 1 \quad & \quad 1 & \quad 2\quad\\
\vspace{-.2cm} &&& \\ \hline \vspace{-.2cm} &&& \\
$n=5$ \quad & \quad 2 \quad & \quad 2 & \quad 1\quad\\
($n=6$) \quad & \quad 2 \quad & \quad 1 & \quad 2\quad\\
$n=7$ \quad & \quad 1 \quad & \quad 2 & \quad 2\quad\\
\vspace{-.2cm} &&& \\ \hline \vspace{-.2cm} &&& \\
$n=8$ \quad & \quad 2 \quad & \quad 2 & \quad 2\quad
\end{tabular}
\label{tabint}
\end{center}
\caption{\label{tab:powers}Overview of the notation for the possible configurations of 
powers of propagators in the $H$ functions in 
PQChPT. Redundant configurations are given in parentheses.}
\end{table}

The interchange $(r,m_1^2,n_1)\leftrightarrow (s,m_2^2,n_2)$
allows to show that $H^s_\mu, H^{ss}_{\mu\nu}$ are related directly to
$H^r_\mu,H^{rr}_{\mu\nu}$. $H^{rs}_{\mu\nu}$ can also be related to
$H_{\mu\nu}$ using the trick shown in \cite{Amoros:1999dp} and also used
in \cite{Bijnens:2013doa}, now taking the pole configurations into account
properly. The resulting $H_{\mu\nu}$ and $H_\mu$ can then be reduced to six
pole configurations only, cf. table \ref{tab:powers}, the bracketed ones can
be eliminated via the interchange above. In the scalar case $H$, only four
pole configurations are needed.

For the partially quenched calculation we thus generalized the sunset
integrals used in our earlier work via
\begin{equation}
H(\chi_i,\chi_j,\chi_k;p^2) \rightarrow 
H(n,\chi_i,\chi_j,\chi_k;p^2),
\end{equation} 
introducing the new index $n$ for the pole configurations as the
first argument. Note on the side that all new pole configurations are related
to the simplest one by differentiation with respect to the mass scales.

In the cms frame, we reduce the tensor structure of the sunsets as
\ba
\label{defHi}
H_\mu &=& p_\mu H_1\,
\\\nonumber
H_{\mu\nu} &=& p_\mu p_\nu H_{21} + g_{\mu\nu} H_{22} + t_{\mu\nu} H_{27}\,.
\ea
As in \cite{Bijnens:2014dea}, we renormalize the FV sunsets by not only
subtracting the infinite part but also an additional finite part
containing ${\cal O}(\epsilon)$ contributions of one-loop integrals.
In this way, the latter integrals will cancel out of the final result, and
thus do not need to be computed. The splitting for $n = 1$
\ba
\tilde H^V &=& \frac{\lambda_0}{16\pi^2}
 \left(A^V(m_1^2)+A^V(m_2^2)+A^V(m_3^2)\right)
 +\frac{1}{16\pi^2}
    \left(A^{V\epsilon}(m_1^2)+A^{V\epsilon}(m_2^2)+A^{V\epsilon}(m_3^2)\right)
\nonumber\\&&
 +H^V\,,
\nonumber\\
\tilde H^V_1 &=& \frac{\lambda_0}{16\pi^2}\frac{1}{2}
 \left(A^V(m_2^2)+A^V(m_3^2)\right)
 +\frac{1}{16\pi^2}\frac{1}{2}
    \left(A^{V\epsilon}(m_2^2)+A^{V\epsilon}(m_3^2)\right)
 +H^V_1\,,
\nonumber\\
\tilde H^V_{21} &=& \frac{\lambda_0}{16\pi^2}\frac{1}{3}
 \left(A^V(m_2^2)+A^V(m_3^2)\right)
 +\frac{1}{16\pi^2}\frac{1}{3}
    \left(A^{V\epsilon}(m_2^2)+A^{V\epsilon}(m_3^2)\right)
 +H^V_{21}\,,
\nonumber\\
\tilde H^V_{27} &=& \frac{\lambda_0}{16\pi^2}
 \left(A^V_{23}(m_1^2)+\frac{1}{3}A_{23}(m_2^2)+\frac{1}{3}A^V_{23}(m_3^2)\right)
\nonumber\\&&
 +\frac{1}{16\pi^2}
    \left(A^{V\epsilon}_{23}(m_1^2)
+\frac{1}{3}A^{V\epsilon}_{23}(m_2^2)+\frac{1}{3}A^{V\epsilon}_{23}(m_3^2)\right)
 +H^V_{27}\,,
\ea
has to be generalized for the other pole configurations by taking the appropriate
derivatives w.r.t. the masses.

\section{Analytical results}
\label{sec:analytical}

\begin{figure}
\begin{center}
\includegraphics[width=7cm]{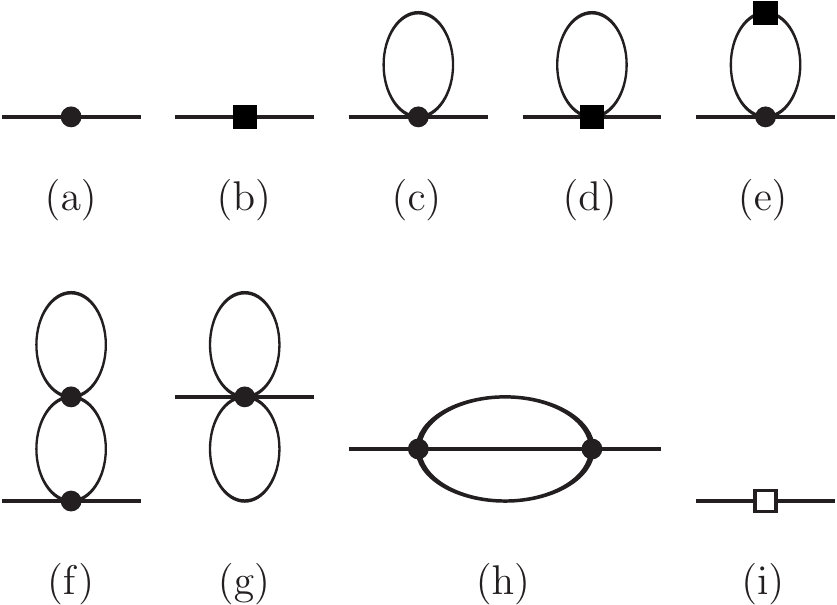}
\caption{\label{massp6diag}
Diagrammatic contributions to the pseudoscalar self-energy,
up to ${\cal O} (p^6)$. Circular vertices are of ${\cal O} (p^2)$,
the filled boxes are of ${\cal O} (p^4)$, the open box is of ${\cal O} (p^6)$.
The tree level diagrams (a,b,i) do not contribute to finite volume
corrections.
}
\end{center}
\end{figure}

The calculation of the masses proceeds in the usual way
from the Feynman diagrams for the self-energy shown in Fig.~\ref{massp6diag}.
We have performed the calculation for the off-diagonal mesons, i.e.
consistening of a valence quark and a different valence anti-quark,
and for the case of three flavours of sea quarks.
The calculation has been done for all mass cases, equal and different
valence quark-masses, $d_{val}=1,2$, and sea quark masses all equal,
$d_{sea}=1$, two equal and the third different, $d_{sea}=2$ and all three
different, $d_{sea}=3$.

A large number of checks have been done on the calculations. They have been
performed both in the supersymmetric formalism and using quark flow techniques.
The infinite volume results are also in full agreement with
\cite{Bijnens:2004hk,Bijnens:2005ae,Bijnens:2006jv}. The finite volume
parts agree with our earlier results \cite{Bijnens:2014dea} when these are
expressed in terms of lowest order masses and when the sea masses are put
equal to the valence masses.

The formulas especially for the case of three different sea quark masses are
very long. In App.~\ref{appmass} we list the case of equal valence masses
and two sea quark masses. This corresponds to the charged pion mass
in the isospin limit. The other cases can be downloaded from \cite{chptweb}.

The masses are given as
\be
m^2_{ij} = \chi_{ij}+m^{2(4)}_{ij}+\Delta^Vm^{2(4)}_{ij}
+m^{2(6)}_{ij}+\Delta^Vm^{2(6)}_{ij}\,.
\ee
In addition a superscript indicating $d_{val}d_{sea}$ is added.
The infinite volume and the one-loop finite volume
corrections were known before. The new parts are the two-loop finite volume
corrections.
These we split in addition in an $L_i^r$ dependent part and a pure
two-loop contribution
\be
\Delta^Vm^{2(6)}_{ij} = \Delta^Vm^{2(6L)}_{ij}+\Delta^Vm^{2(6R)}_{ij}\,.
\ee
The subscript $ij$ is set to 12 for $d_{val}=1$ and to 13 for $d_{val}=2$
similar to the infinite volume work.
 
The decay constant is defined in the usual way as
\be
\langle0|\bar q_j\gamma_\mu\gamma_5 q_i | M_{ij}(p)\rangle
= i\sqrt{2} F_{ij} p_\mu\,,
\ee
for the pseudoscalar meson $M_{ij}$ with quark content $i\ne j$ and momentum
$p$.
The calculation needs the the diagrams of Fig.~\ref{massp6diag} for the
wave function renormalization and the same ones with one external meson leg
replaced by an insertion of the axial current.

We split the result as
\be
F_{ij} = F_0+F^{(4)}_{ij}+\Delta^VF^{(4)}_{ij}
+F^{(6)}_{ij}+\Delta^VF^{(6)}_{ij}\,.
\ee
The NNLO part is split again in
\be
\Delta^VF^{(6)}_{ij} = \Delta^VF^{(6L)}_{ij}+\Delta^VF^{2(6R)}_{ij}\,.
\ee

The calculations have been done using the supersymmetric and the
quark flow methods. The infinite volume and NLO results agree
with the known expressions and the result reduces in the correct limit
to the unquenched results of our earlier work \cite{Bijnens:2014dea}.
The formulas are rather long, the case for equal valence masses and two
different sea masses corresponding to the charged pion decay constant
in the isospin limit is given in App.~\ref{appdecay}. The expressions
for the other cases can be downloaded from \cite{chptweb}.

\section{Numerical examples}
\label{sec:numerical}

The intention is that various lattice QCD collaborations can use our formulas.
All cases discussed have been included in the package
\textsc{CHIRON} \cite{Bijnens:2014gsa} available from \cite{chiron}.
The numerical results shown in this section have been obtained with that
implementation.
The programs have been cross-checked with an independent version.
It has been checked that the results reduce in the appropriate limits
to those of our earlier work \cite{Bijnens:2014dea}. For this purpose
the expressions obtained in \cite{Bijnens:2014dea}, but rewritten in terms of
lowest order masses and decay constants, have been implemented and
included in \textsc{CHIRON} \cite{chiron}. In addition,
a check has been done that the different mass cases reduce to each other
numerically.

For input values we have chosen the recent global fit for
the $L_i^r$ \cite{Bijnens:2014lea}. We have set the extra LEC $L_0^r=0$.
We always use a scale of $\mu=0.77$~GeV.
For the size of the lattice we present results for a length $L$ such
that $M L=2$ for $M=0.13$~GeV. The lowest order pion decay constant we have
chosen throughout as $F_0=87.7$~MeV.

The numerical results are presented via
\ba
\label{defDV}
\Delta^V_M &=& \frac{m^{2V}_{ij}-m^{2\infty}_{ij}}{\chi_{ij}}\,
\nonumber\\
\Delta^V_F &=& \frac{F_{ij}^V-F_{ij}^\infty}{F_0}\,.
\ea
We thus plot the size of the finite volume corrections relative to the
lowest order value of the quantity under consideration.
Note that the results are for charged or off-diagonal mesons. They
consist of a quark and a different anti-quark which might have equal
mass.

\subsection{\boldmath $d_\mathrm{val}=d_\mathrm{sea}=1$}

\begin{figure}
\includegraphics[width=0.49\textwidth]{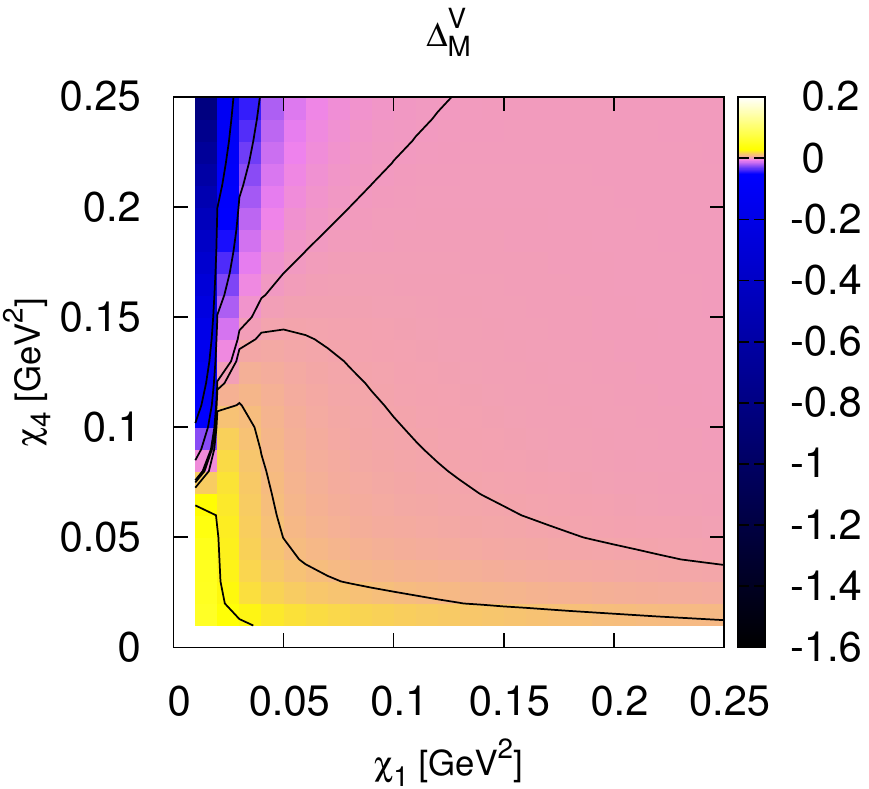}
\includegraphics[width=0.49\textwidth]{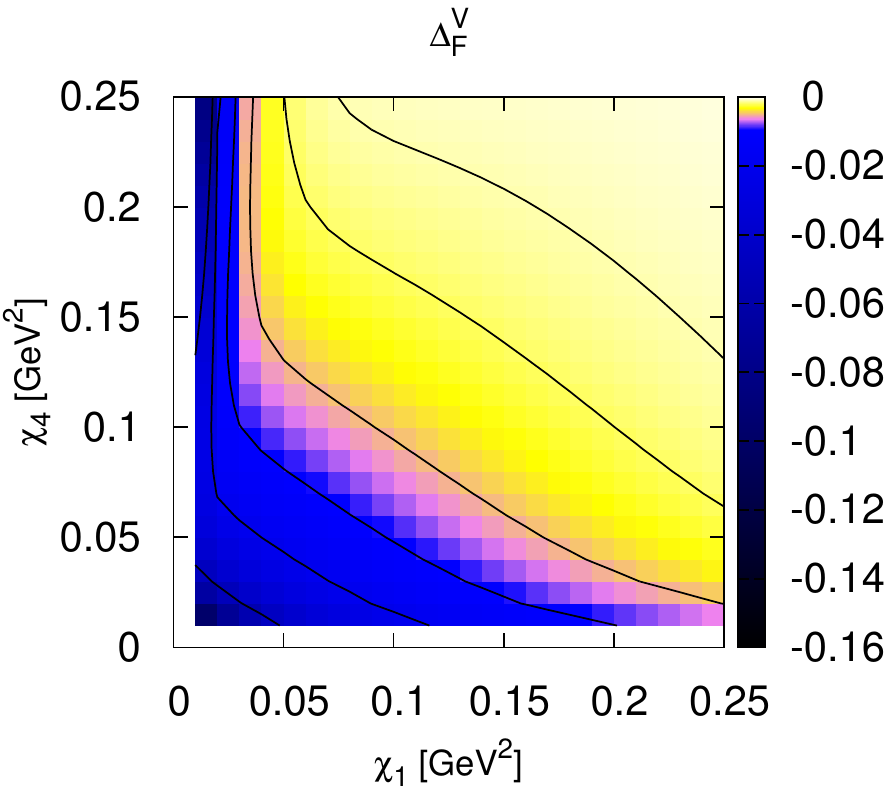}
\caption{\label{fig11} The finite volume corrections relative to the
lowest order value as defined in (\ref{defDV}) for the case
with all valence masses equal and all sea masses equal.
Left: $\Delta^V_M$ the correction to the mass-squared,
contour lines are drawn at  $0.03,0.01,0.003,0,-0.03,-0.01$ starting from the
bottom left and going counterclockwise. Right: The correction
to the decay constant, contour lines are drawn at $ -0.001,-0.002,-0.005,-0.01,-0.02,-0.05$ going from top-right to bottom-left.}
\end{figure}

Here we set all valence and all sea masses equal,
$d_\mathrm{val}=d_\mathrm{sea}=1$. The size of the finite volume corrections
as a function of $\chi_1$ and $\chi_4$ is shown in Fig.~\ref{fig11}.
The corrections in this case are reasonable, at most a few \%,
except for very low masses and become very large for low valence and high sea
quark mass.

\subsection{The pion mass}

In this subsection we look at the case where the lowest order mass
is around the pion mass. We plot $\Delta^V_M$ with $\sqrt{\chi_{12}}=0.13$~GeV.
The strange sea quark mass we have always chosen such that the average
lowest order kaon mass is $0.45$~GeV. This corresponds to
$\sqrt{\chi_6} = \sqrt{2(0.45)^2-(0.13)^2}$~GeV $\approx 0.623$~GeV.
The other input parameters are chosen as given in the introduction
of this section. We have restricted the sea up and down quark masses
corresponding to a lowest order sea quark pion of 100 to 300~MeV.

\begin{figure}
\begin{minipage}{0.49\textwidth}
\includegraphics[width=0.99\textwidth]{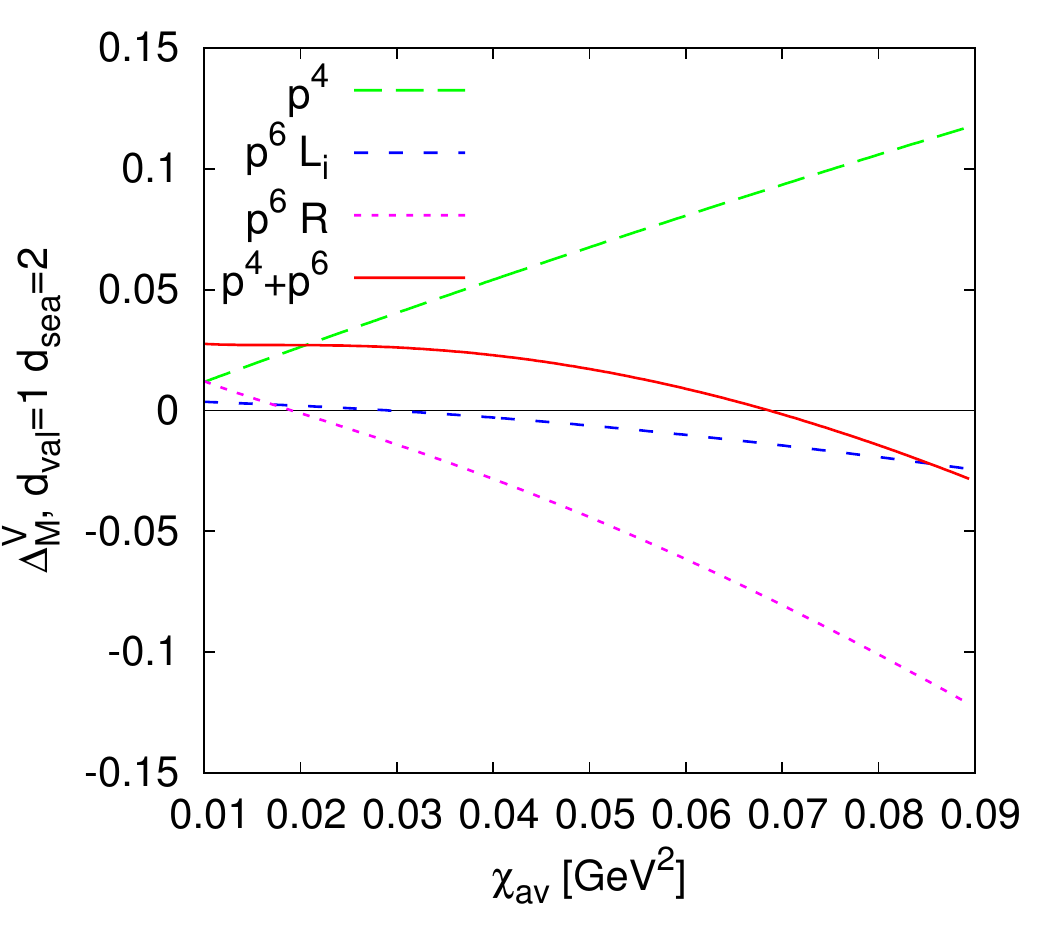}
\centerline{(a)}
\end{minipage}
\begin{minipage}{0.49\textwidth}
\includegraphics[width=0.99\textwidth]{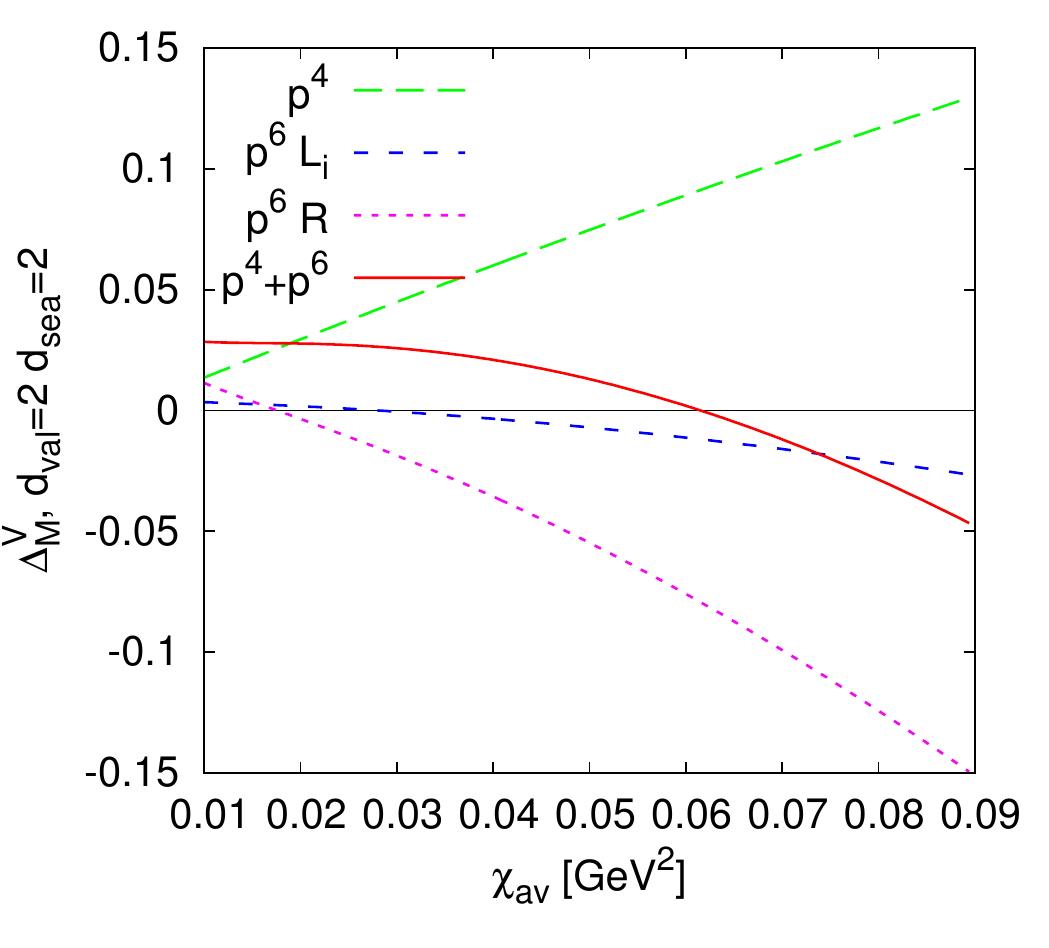}
\centerline{(b)}
\end{minipage}
\begin{minipage}{0.49\textwidth}
\includegraphics[width=0.99\textwidth]{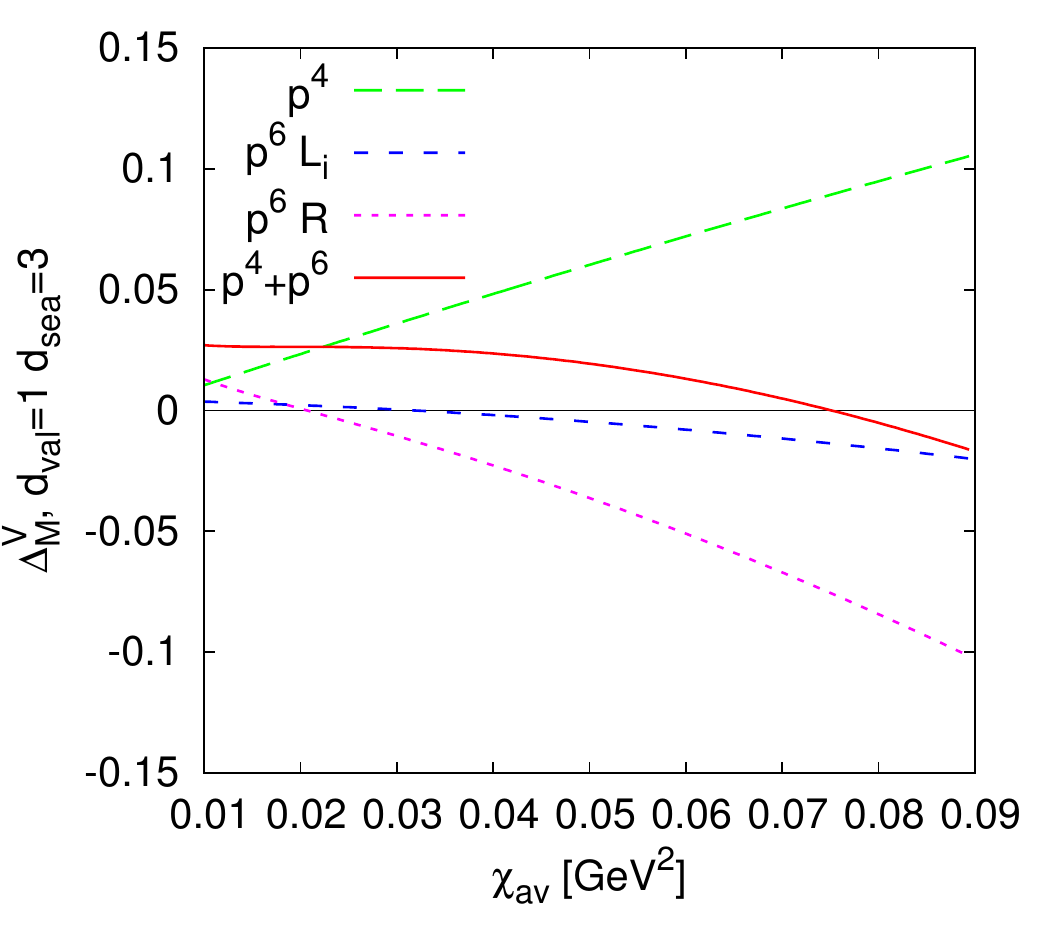}
\centerline{(c)}
\end{minipage}
\begin{minipage}{0.49\textwidth}
\includegraphics[width=0.99\textwidth]{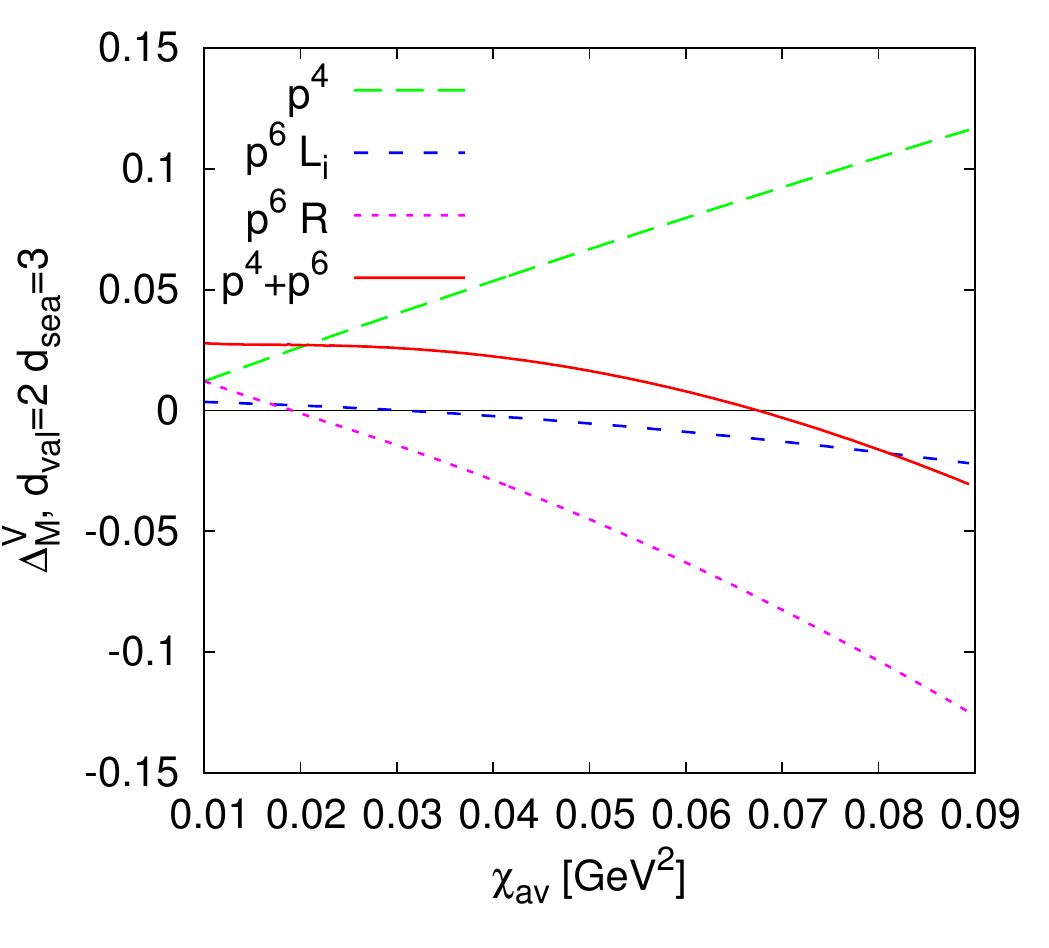}
\centerline{(d)}
\end{minipage}
\caption{\label{figmpi} The corrections for the pion mass relative
to the lowest order mass as a function
of the average up and down sea quark mass via $\chi_\mathrm{av}$.
(a) The isospin limit, $\chi_1=\chi_2$, $\chi_4=\chi_5=\chi_\mathrm{av}$.
(b) Isospin breaking in the valence sector, $\chi_1=\chi_3/2$ and
$\chi_4=\chi_5=\chi_\mathrm{av}$.
(c) Isospin breaking in the sea sector, $\chi_1=\chi_2$ and
$\chi_4=\chi_5/2$.
(d) Isospin breaking in both sectors, $\chi_1=\chi_3/2$ and
$\chi_4=\chi_5/2$.}
\end{figure}

The first case we look at is $d_\mathrm{val}=1,d_\mathrm{sea}=2$. This
corresponds to taking the up and down quark masses equal in both
the valence and sea quark sector and a different
strange quark mass. This is the isospin limit.
The result is shown in Fig.~\ref{figmpi}(a).
There is a rather large cancellation between the
$p^4$ and $p^6$ correction while the $p^6$ contribution coming from the $L_i^r$
is fairly small.

We now include isospin breaking in the valence sector. We thus look at the
case with $d_\mathrm{val}=2, d_\mathrm{sea}=2$.
We fix the valence quark masses such that $\chi_1+\chi_2=2\chi_{12}$
and $\chi_1/\chi_2=1/2$. There is a sizable isospin breaking visible
in the finite volume corrections, as shown in Fig.~\ref{figmpi}(b).

The opposite case, isospin breaking in the sea sector, but not in the valence
sector, leads to numerically similar but opposite sign corrections.
Here we used $\chi_1=\chi_2$, $\chi_4=\chi_5/2$ and 
$\chi_4+\chi_5=2\chi_\mathrm{av}$. The results are shown
in Fig.~\ref{figmpi}(c).

Finally, we introduce isospin breaking in both the valence and sea quark sector
with  $\chi_1/\chi_2=1/2$, $\chi_4=\chi_5/2$ and 
$\chi_4+\chi_5=2\chi_\mathrm{av}$. The results are shown
in Fig.~\ref{figmpi}(d). The total isospin corrections are rather small.

\begin{figure}
\begin{center}
\includegraphics[width=0.49\textwidth]{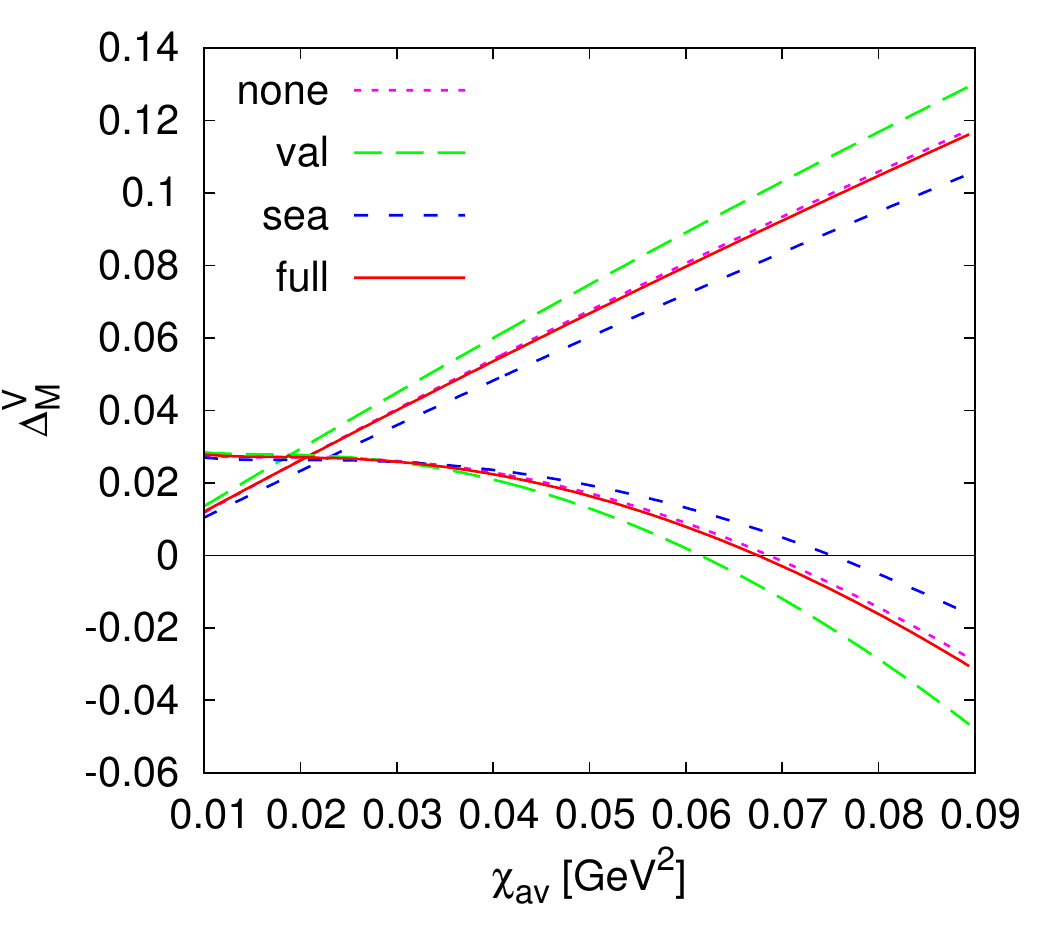}
\end{center}
\caption{\label{figiso} Comparing the finite volume correction for
the meson masses for the cases with no isospin breaking (none),
only in the valence sector (val), only in the sea sector (sea) and in both
(full) for the meson mass squared. The upper curves are the $p^4$, the bottom the $p^4+p^6$ results.}
\end{figure}

The numerical cancellation between the isospin breaking in the
valence and sea quark case is accidental. The corrections due to
valence and sea quark masses are all second order in isopin breaking.
The same argument as in the unquenched case goes through both for the
valence and sea quark masses. We have compared four scenarios in
Fig.~\ref{figiso}. We show the $p^4$ and the full $p^4+p^6$ result
first with no isospin breaking, then
only in the valence sector or only in the sea sector
and finally in both sectors. The curves are those shown in
Fig.~\ref{figmpi}(a-d). We have checked numerically by using a different
ratio for the isospin breaking that the corrections are indeed second order
in isospin breaking.

\subsection{The pion decay constant}

In this subsection we look at the same cases
as before. The lowest order mass
is around the pion mass. We plot $\Delta^V_F$ with $\sqrt{\chi_{12}}=0.13$~GeV.
and as before $\sqrt{\chi_6} = \sqrt{2(0.45)^2-(0.13)^2}$~GeV $\approx 0.623$~GeV.
The other input parameters are again chosen as given in introduction
of this section. We have restricted the sea up and down quark masses
corresponding to a lowest order sea quark pion of 100 to 300~MeV.

\begin{figure}
\begin{minipage}{0.49\textwidth}
\includegraphics[width=0.99\textwidth]{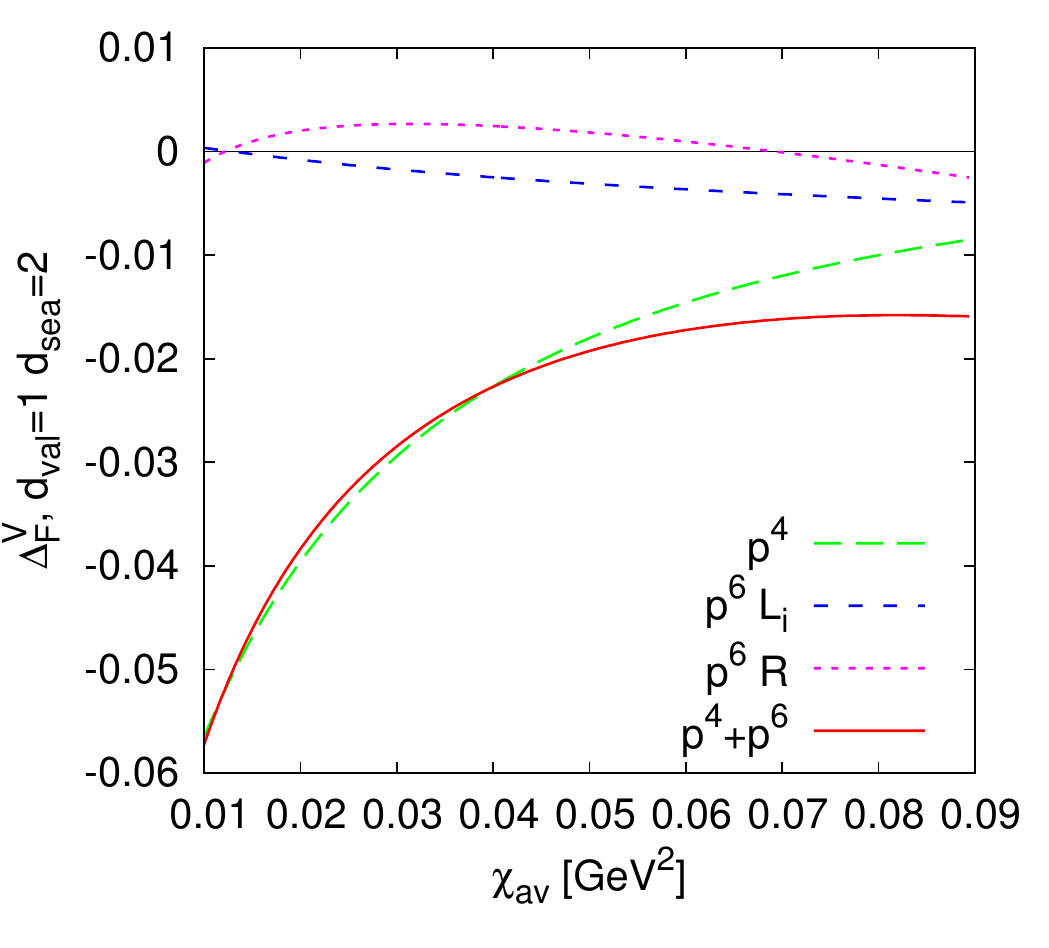}
\centerline{(a)}
\end{minipage}
\begin{minipage}{0.49\textwidth}
\includegraphics[width=0.99\textwidth]{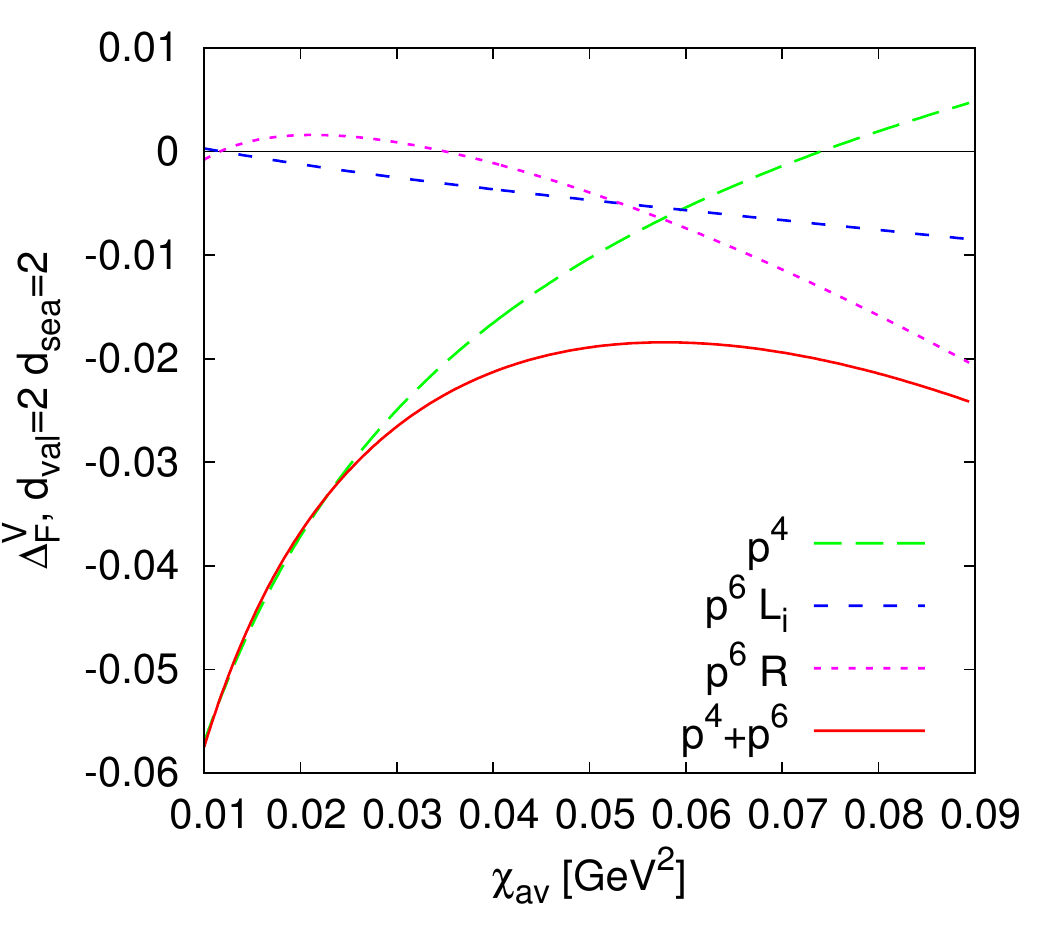}
\centerline{(b)}
\end{minipage}
\begin{minipage}{0.49\textwidth}
\includegraphics[width=0.99\textwidth]{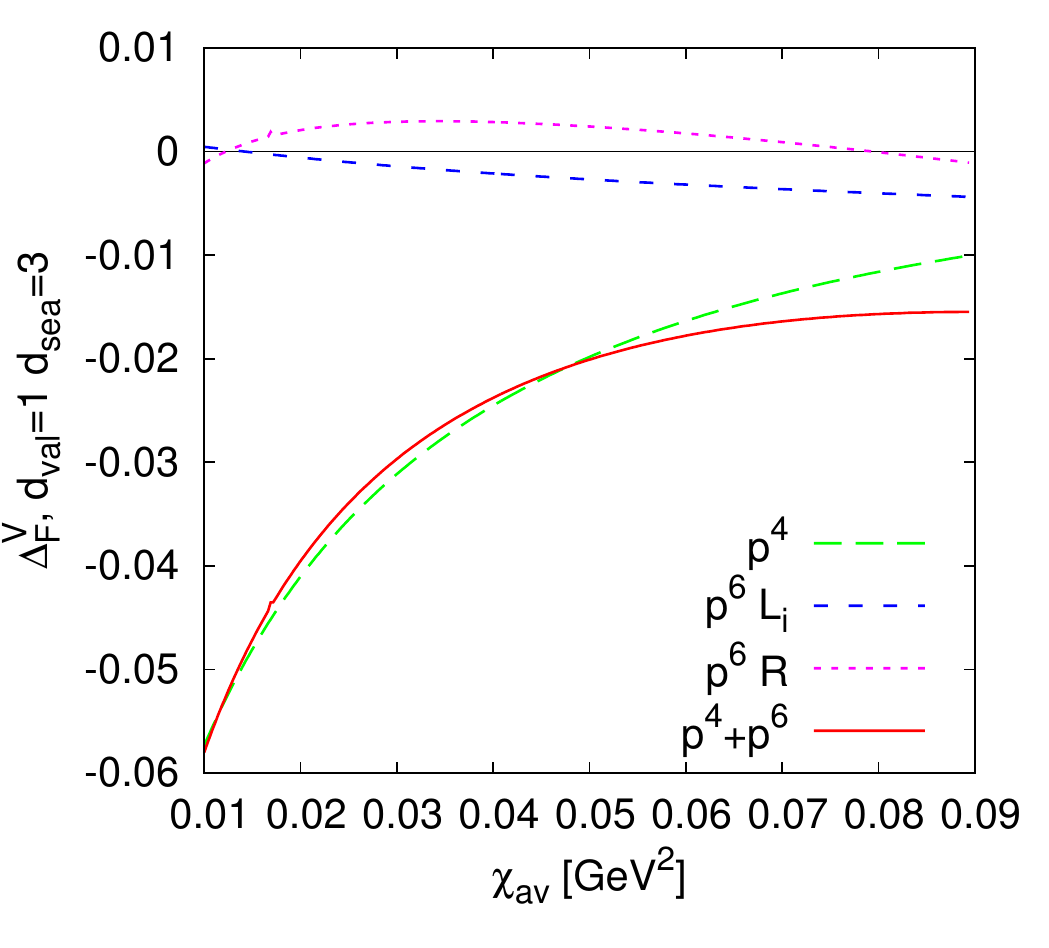}
\centerline{(c)}
\end{minipage}
\begin{minipage}{0.49\textwidth}
\includegraphics[width=0.99\textwidth]{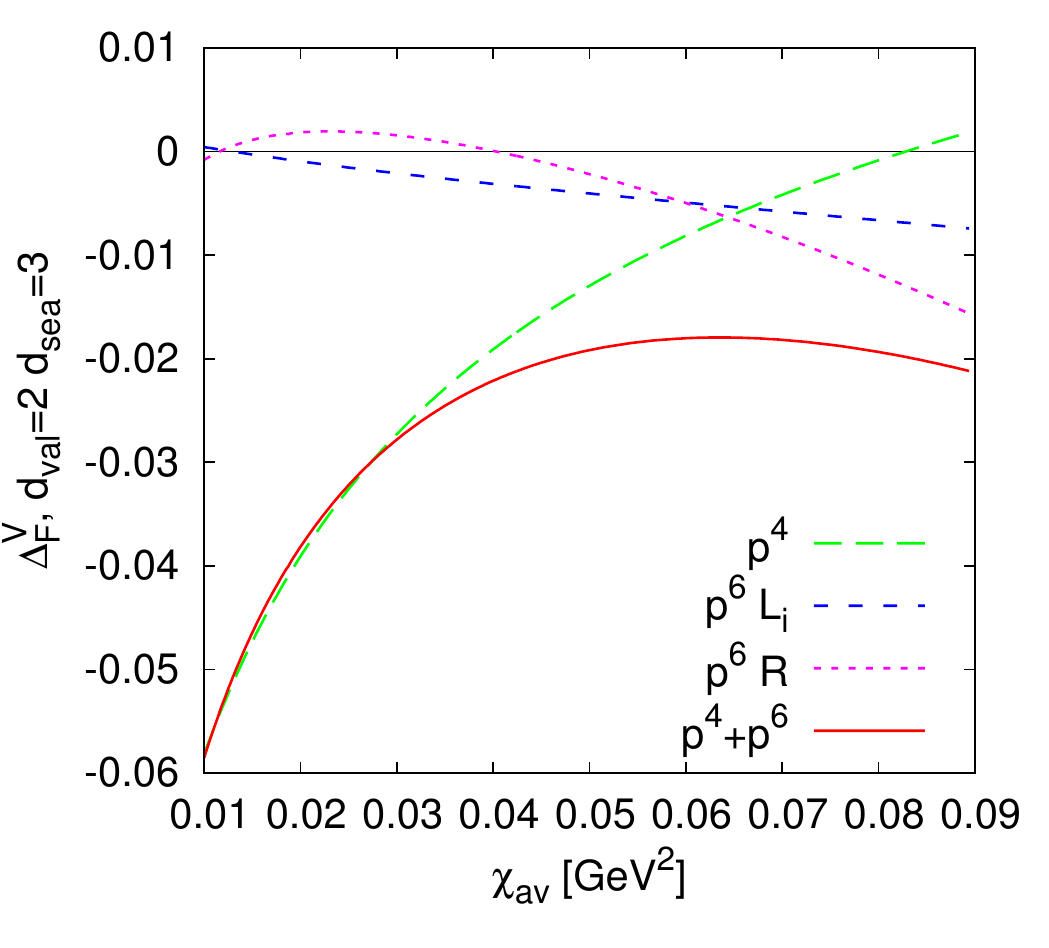}
\centerline{(d)}
\end{minipage}
\caption{\label{figfpi} The corrections for the pion decay constant relative
to its lowest order value as a function
of the average up and down sea quark mass via $\chi_\mathrm{av}$.
(a) The isospin limit, $\chi_1=\chi_2$, $\chi_4=\chi_5=\chi_\mathrm{av}$.
(b) Isospin breaking in the valence sector, $\chi_1=\chi_3/2$ and
$\chi_4=\chi_5=\chi_\mathrm{av}$.
(c) Isospin breaking in the sea sector, $\chi_1=\chi_2$ and
$\chi_4=\chi_5/2$.
(d) Isospin breaking in both sectors, $\chi_1=\chi_3/2$ and
$\chi_4=\chi_5/2$.}
\end{figure}

The first case we look at is $d_\mathrm{val}=1,d_\mathrm{sea}=2$. This
corresponds to taking the up and down quark masses equal in both
the valence and sea quark sector and a different
strange quark mass, i.e. the isospin limit.
The result is shown in Fig.~\ref{figfpi}(a).
The total $p^6$ correction is fairly small.

We now include isospin breaking in the valence sector. We thus look at the
case with $d_\mathrm{val}=2, d_\mathrm{sea}=2$.
We fix the valence quark masses such that $\chi_1+\chi_2=2\chi_{12}$
and $\chi_1/\chi_2=1/2$. There is a sizable isospin breaking visible
in the finite volume corrections, as shown in Fig.~\ref{figmpi}(b).

The opposite case, isospin breaking in the sea sector but not in the valence
sector leads to numerically much smaller effects.
Here we used $\chi_1=\chi_2$, $\chi_4=\chi_5/2$ and 
$\chi_4+\chi_5=2\chi_\mathrm{av}$. The results are shown
in Fig.~\ref{figmpi}(c).

Finally, we introduce isospin breaking in both the valence and sea quark sector
with  $\chi_1/\chi_2=1/2$, $\chi_4=\chi_5/2$ and 
$\chi_4+\chi_5=2\chi_\mathrm{av}$. The results are shown
in Fig.~\ref{figmpi}(d). The total isospin corrections are failry small.

The corrections due to
valence and sea quark masses are all second order in isospin breaking.
The same argument as in the unquenched case goes through both for the
valence and sea quark masses. 
We compare the same four scenarios as for the pion mass,
no isospin breaking,
only in the valence sector, only in the sea sector and in both sectors.
The curves are those shown in
Fig.~\ref{figfpi}(a-d). 
\begin{figure}
\begin{minipage}{0.49\textwidth}
\includegraphics[width=0.99\textwidth]{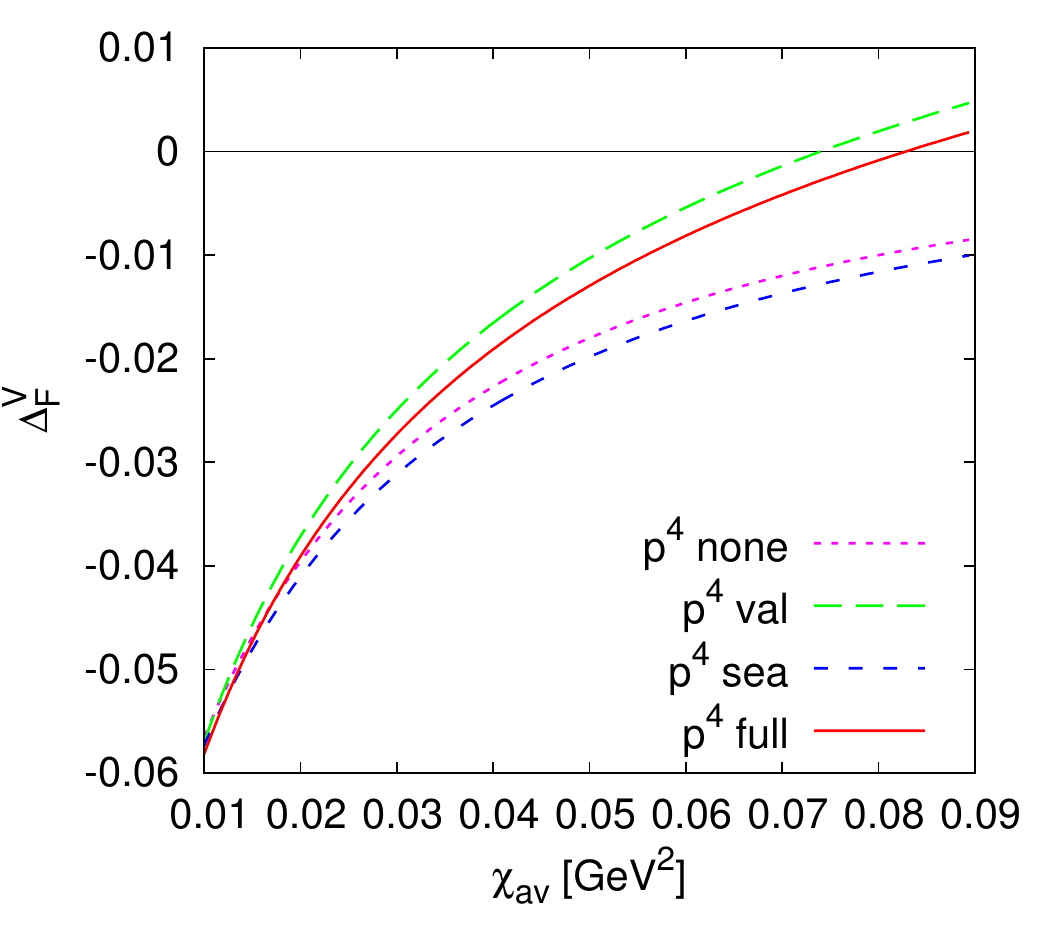}
\centerline{(a)}
\end{minipage}
\begin{minipage}{0.49\textwidth}
\includegraphics[width=0.99\textwidth]{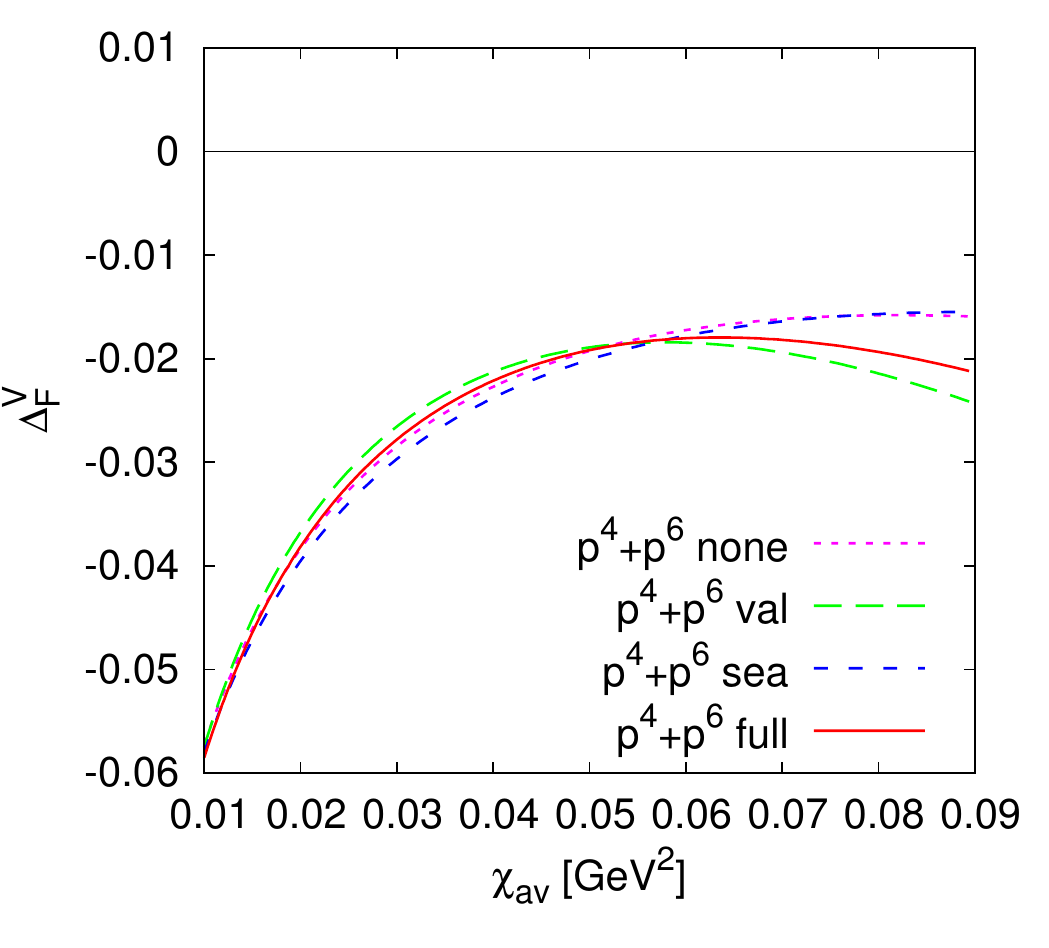}
\centerline{(b)}
\end{minipage}
\caption{\label{figfpiiso} Comparing the finite volume correction for
the meson decay constant and masses for the cases with no isospin
breaking (none),
only in the valence sector (val), only in the sea sector (sea) and in both
(full) for the meson mass squared.
(a) $p^4$ (b) $p^4+p^6$.}
\end{figure}

\subsection{The kaon mass and decay constant}

\begin{figure}
\begin{minipage}{0.49\textwidth}
\includegraphics[width=0.99\textwidth]{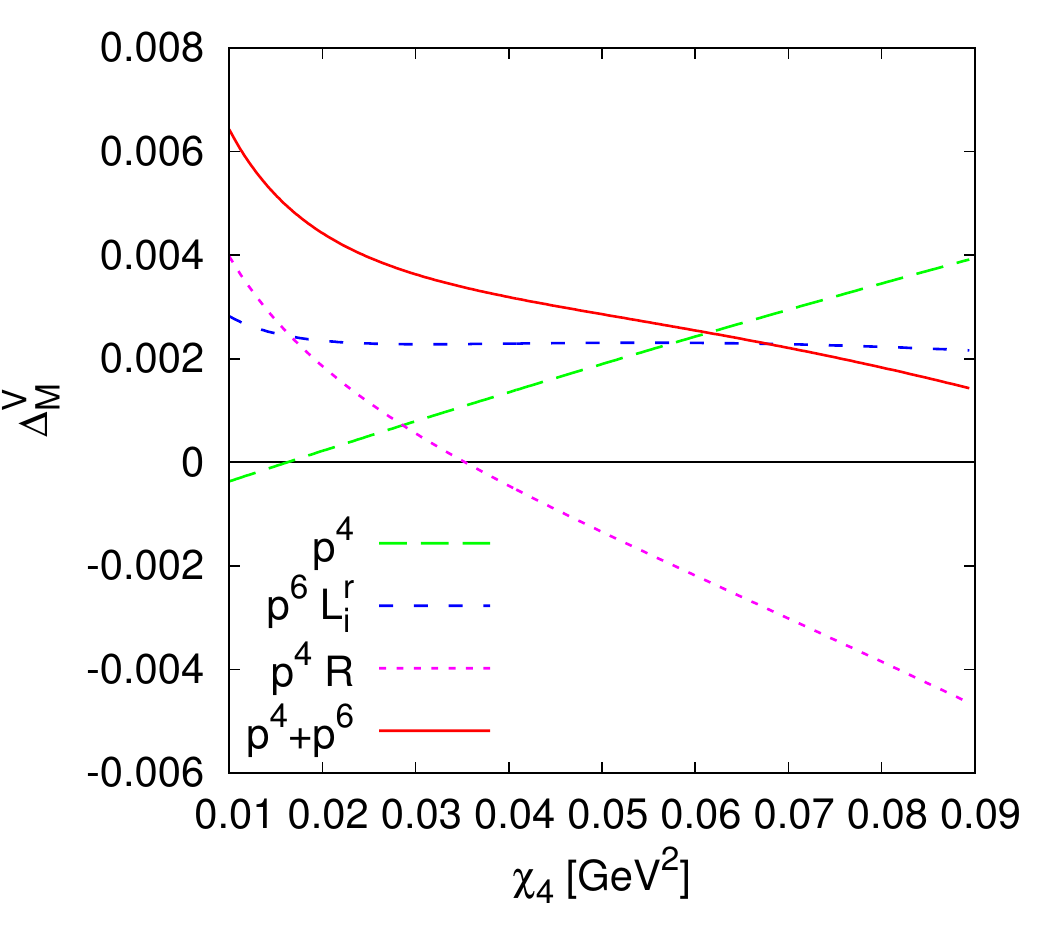}
\centerline{(a)}
\end{minipage}
\begin{minipage}{0.49\textwidth}
\includegraphics[width=0.99\textwidth]{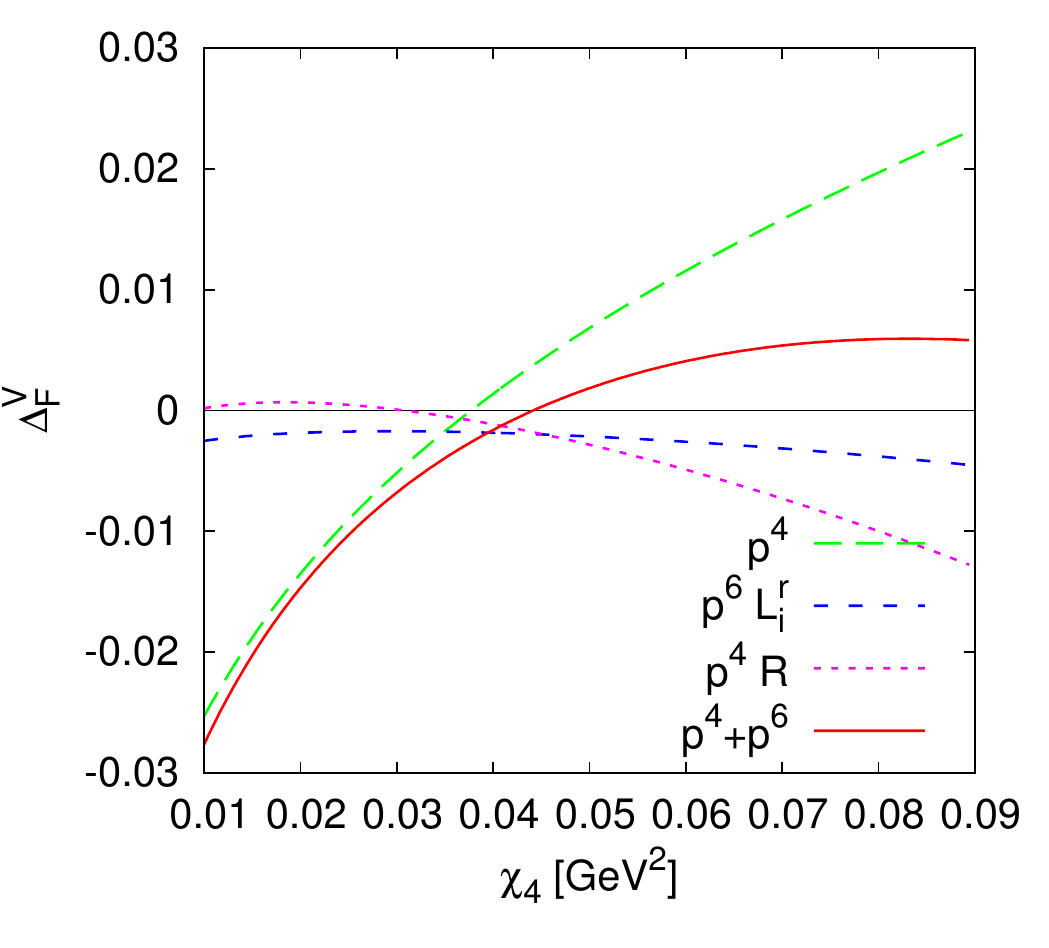}
\centerline{(b)}
\end{minipage}
\caption{\label{figkaon}
The finite volume corrections for a valence mass close to the kaon mass
relative to the lowest order value.
(a) the kaon mass squared. (b) the kaon decay constant.}
\end{figure}

We now look only at the $d_\mathrm{val}=d_\mathrm{sea}=2$ case
but choose the valence masses such that we have a lowest order pion mass
of 130~MeV and a lowest order kaon mass of 450~MeV.
This corresponds to $\sqrt{\chi_1}=130$~MeV and $\sqrt{\chi_3}\approx623$~MeV.
We plot the finite volume corrections relative to the lowest order value
of the quantity in Fig.~\ref{figkaon} as a function of $\chi_4=\chi_5$.
For the sea quark strange mass we  use $\chi_6=1.02\chi_3$.
The LECs are again the ones from \cite{Bijnens:2014lea}
and $L$ such that $ML=2$ for $M=$130~MeV. 

For the kaon we see that we reproduce the results of \cite{Bijnens:2014dea}
that near the physical case the $p^4$ corrections are very small.
The total finite volume corrections to the mass remain fairly small.
The kaon decay constant has larger corrections but they remain in the
few \% region for the parameters considered.

\section{Conclusions}

We have computed the NNLO expressions for the masses and decay constants
in three-flavour partially quenched ChPT for all possible mass cases.
The calculation has been performed using two different formalisms, quark-flow
and the supersymmetric method. The known infinite volume expressions
have been reproduced.
We quoted the expressions for the equal valence and two different sea quark
masses in the appendices. The other cases can be obtained from
\cite{chptweb}.

The numerical work shows finite volume corrections of a similar
size as those in the unquenched case \cite{Bijnens:2014dea}. We have
presented some representative numerics.
The numerical work has been done using \textsc{C++}. The programs are
available together with the infinite volume results in \cite{chiron}.
The analytical work relied heavily on \textsc{FORM} \cite{FORM}.

\section*{Acknowledgements}

This work is supported in part by the Swedish Research Council grants
621-2011-5080 and 621-2013-4287. JB thanks the Centro de Ciencias
de Benasque Pedro Pascual, where part of this paper was written,
for hospitality.

\appendix
\section{Expressions for the mass}
\label{appmass}

\begin{equation}
F_0^2 \Delta^V\! m^{2(4)12}_{12} =
       + {A}^{V}(\chi_{1}) \, \Big(  - 1/3\,\chi_{1}\,{R^c_{146\eta}}  \Big)
       + {A}^{V}(\chi_\eta) \, \Big( 2/3\,\chi_{1}\,{R^z_{\eta 61}} ^2 \Big)
       + {B}^{V}(\chi_{1}) \, \Big(  - 1/3\,\chi_{1}\,{R^z_{146\eta}}  \Big)
\end{equation}
\begin{eqnarray}
\lefteqn{F_0^4 \Delta^V\! m^{2(6L)12}_{12} =
       + {A}^{V}(\chi_{1}) \, \Big(  - 64/3\,\hat L^r_{8}\chi_{1}^2\,{R^c_{146\eta}}  - 32/3\,\hat L^r_{7}\chi_{1}\,\chi_{6}\,{R^z_{14\eta}}  - 64/3\,\hat L^r_{7}\chi_{1}\,\chi_{4}\,{R^z_{16\eta}}
}&&
\nonumber\\&&
       + 32\,\hat L^r_{7}\chi_{1}^2 + 32\,\hat L^r_{6}\chi_{1}^2 + 8/3\,\hat L^r_{5}\chi_{1}\,{R^z_{146\eta}}  - 8/9\,\hat L^r_{5}\chi_{1}\,\chi_{6}\,{R^z_{14\eta}} ^2 - 16/9\,\hat L^r_{5}\chi_{1}\,
         \chi_{4}\,{R^z_{16\eta}} ^2
\nonumber\\&&
 + 40/3\,\hat L^r_{5}\chi_{1}^2\,{R^c_{146\eta}}  - 8\,\hat L^r_{3}\chi_{1}\,{R^z_{146\eta}}  - 8\,\hat L^r_{3}\chi_{1}^2\,{R^c_{146\eta}}  + 20\,\hat L^r_{2}\chi_{1}^2
          + 8\,\hat L^r_{1}\chi_{1}^2 - 8\,\hat L^r_{0}\chi_{1}\,{R^z_{146\eta}} 
\nonumber\\&&
 - 8\,\hat L^r_{0}\chi_{1}^2\,{R^c_{146\eta}}  - 16/3\,(\chi_{6} + 2\,\chi_{4})\,\hat L^r_{6}\chi_{1}\,
         {R^c_{146\eta}}  - 16/9\,(\chi_{6} + 2\,\chi_{4})\,\hat L^r_{4}\chi_{1}\,{R^z_{16\eta}} ^2 
\nonumber\\&&
- 8/9\,(\chi_{6} + 2\,\chi_{4})\,\hat L^r_{4}\chi_{1}\,{R^z_{14\eta}} ^2 + 8\,
         \Big(\chi_{6} + 2\,\chi_{4})\,\hat L^r_{4}\chi_{1}\,{R^c_{146\eta}}  \Big)
       + {A}^{V}(\chi_{14}) \, \Big( 32\,(\chi_{4} + \chi_{1})\,\hat L^r_{8}\chi_{1}
\nonumber\\&&
 - 16\,(\chi_{4} + \chi_{1})\,\hat L^r_{5}\chi_{1} + 20\,(\chi_{4} + \chi_{1})\,\hat L^r_{3}
         \chi_{1} + 8\,(\chi_{4} + \chi_{1})\,\hat L^r_{0}\chi_{1} \Big)
       + {A}^{V}(\chi_{16}) \, \Big( 16\,(\chi_{6} + \chi_{1})\,\hat L^r_{8}\chi_{1}
\nonumber\\&&
 - 8\,(\chi_{6} + \chi_{1})\,\hat L^r_{5}\chi_{1} + 10\,(\chi_{6} + \chi_{1})\,\hat L^r_{3}\chi_{1}
          + 4\,(\chi_{6} + \chi_{1})\,\hat L^r_{0}\chi_{1} \Big)
       + {A}^{V}(\chi_{4}) \, \Big( 48\,\hat L^r_{6}\chi_{1}\,\chi_{4} 
\nonumber\\&&
- 48\,\hat L^r_{4}\chi_{1}\,\chi_{4} + 12\,\hat L^r_{2}\chi_{1}\,\chi_{4} + 48\,\hat L^r_{1}\chi_{1}\,\chi_{4} \Big)
       + {A}^{V}(\chi_{46}) \, \Big( 32\,(\chi_{6} + \chi_{4})\,\hat L^r_{6}\chi_{1} - 32\,(\chi_{6} + \chi_{4})\,\hat L^r_{4}\chi_{1} 
\nonumber\\&&
+ 8\,(\chi_{6} + \chi_{4})\,\hat L^r_{2}\chi_{1}
          + 32\,(\chi_{6} + \chi_{4})\,\hat L^r_{1}\chi_{1} \Big)
       + {A}^{V}(\chi_\eta) \, \Big( 128/3\,\hat L^r_{8}\chi_{1}^2\,{R^z_{\eta 61}} ^2 
\nonumber\\&&
+ 32/9\,\hat L^r_{5}\chi_{1}\,\chi_{6}\,{R^z_{14\eta}} \,{R^z_{\eta 61}}  - 32/9\,\hat L^r_{5}\chi_{1}\,\chi_{4}\,
         {R^z_{16\eta}} \,{R^z_{\eta 61}}  + 16\,\hat L^r_{3}\chi_{1}\,\chi_\eta\,{R^z_{\eta 61}} ^2 + 4\,\hat L^r_{2}\chi_{1}\,\chi_\eta 
\nonumber\\&&
+ 16\,\hat L^r_{1}\chi_{1}\,\chi_\eta + 16\,\hat L^r_{0}\chi_{1}\,\chi_\eta\,
         {R^z_{\eta 61}} ^2 - 8/3\,(3\,\chi_\eta + 2\,\chi_{6} + \chi_{4})\,\hat L^r_{4}\chi_{1} 
\nonumber\\&&
- 16/9\,(3\,\chi_\eta + 2\,\chi_{6} + \chi_{4} + 12\,\chi_{1})\,\hat L^r_{5}
         \chi_{1}\,{R^z_{\eta 61}} ^2 + 64/3\,(\chi_{6} - \chi_{4})\,\hat L^r_{7}\chi_{1}\,{R^z_{\eta 61}}  
\nonumber\\&&
+ 32/3\,(\chi_{6} + 2\,\chi_{4})\,\hat L^r_{6}\chi_{1}\,{R^z_{\eta 61}} ^2 - 64/
         3\,(\chi_{6} + 2\,\chi_{4})\,\hat L^r_{4}\chi_{1}\,{R^z_{\eta 61}} ^2 
\nonumber\\&&
- 32/9\,(\chi_{6} + 2\,\chi_{4})\,\hat L^r_{4}\chi_{1}\,{R^z_{16\eta}} \,{R^z_{\eta 61}}  + 32/9\,(\chi_{6}
          + 2\,\chi_{4})\,\hat L^r_{4}\chi_{1}\,{R^z_{14\eta}} \,{R^z_{\eta 61}}  + 16/3\,(2\,\chi_{6} + \chi_{4})\,\hat L^r_{6}\chi_{1} \Big)
\nonumber\\&&
       + {B}^{V}(\chi_{1}) \, \Big( 16/9\,\hat L^r_{8}\chi_{1}\,\chi_{6}^2\,{R^z_{14\eta}} ^2 + 32/9\,\hat L^r_{8}\chi_{1}\,\chi_{4}^2\,{R^z_{16\eta}} ^2 - 64/3\,\hat L^r_{8}\chi_{1}^2\,
         {R^z_{146\eta}}  - 32/3\,\hat L^r_{8}\chi_{1}^3\,{R^c_{146\eta}}  
\nonumber\\&&
+ 16/9\,\hat L^r_{7}\chi_{1}\,\chi_{6}^2\,{R^z_{14\eta}} ^2 + 64/9\,\hat L^r_{7}\chi_{1}\,\chi_{4}\,\chi_{6}\,{R^z_{14\eta}} \,
         {R^z_{16\eta}}  + 64/9\,\hat L^r_{7}\chi_{1}\,\chi_{4}^2\,{R^z_{16\eta}} ^2 - 32/3\,\hat L^r_{7}\chi_{1}^2\,\chi_{6}\,{R^z_{14\eta}} 
\nonumber\\&&
 - 64/3\,\hat L^r_{7}\chi_{1}^2\,\chi_{4}\,{R^z_{16\eta}}
          + 16\,\hat L^r_{7}\chi_{1}^3 + 40/3\,\hat L^r_{5}\chi_{1}^2\,{R^z_{146\eta}}  - 8/9\,\hat L^r_{5}\chi_{1}^2\,\chi_{6}\,{R^z_{14\eta}} ^2 - 16/9\,\hat L^r_{5}\chi_{1}^2\,\chi_{4}\,
         {R^z_{16\eta}} ^2 
\nonumber\\&&
+ 16/3\,\hat L^r_{5}\chi_{1}^3\,{R^c_{146\eta}}  - 8\,\hat L^r_{3}\chi_{1}^2\,{R^z_{146\eta}}  - 8\,\hat L^r_{0}\chi_{1}^2\,{R^z_{146\eta}}  - 16/3\,(\chi_{6} +
         2\,\chi_{4})\,\hat L^r_{6}\chi_{1}\,{R^z_{146\eta}} 
\nonumber\\&&
 + 16/9\,(\chi_{6} + 2\,\chi_{4})\,\hat L^r_{6}\chi_{1}\,\chi_{6}\,{R^z_{14\eta}} ^2 + 32/9\,(\chi_{6} + 2\,\chi_{4})\,
         \hat L^r_{6}\chi_{1}\,\chi_{4}\,{R^z_{16\eta}} ^2 
\nonumber\\&&
- 32/3\,(\chi_{6} + 2\,\chi_{4})\,\hat L^r_{6}\chi_{1}^2\,{R^c_{146\eta}}  + 8\,(\chi_{6} + 2\,\chi_{4})\,\hat L^r_{4}\chi_{1}\,
         {R^z_{146\eta}}  - 16/9\,(\chi_{6} + 2\,\chi_{4})\,\hat L^r_{4}\chi_{1}^2\,{R^z_{16\eta}} ^2 
\nonumber\\&&
- 8/9\,(\chi_{6} + 2\,\chi_{4})\,\hat L^r_{4}\chi_{1}^2\,{R^z_{14\eta}} ^2 +
         16/3\,(\chi_{6} + 2\,\chi_{4})\,\hat L^r_{4}\chi_{1}^2\,{R^c_{146\eta}}  \Big)
         \nonumber\\&&
       + {B}^{V}(\chi_{1},\chi_\eta) \, \Big(  - 64/9\,\hat L^r_{8}\chi_{1}\,\chi_{6}^2\,{R^z_{14\eta}} \,{R^z_{\eta 61}}  + 64/9\,\hat L^r_{8}\chi_{1}\,\chi_{4}^2\,{R^z_{16\eta}} \,{R^z_{\eta 61}}  + 64/
         3\,\hat L^r_{8}\chi_{1}^3\,{R^z_{\eta 61}} ^2 
         \nonumber\\&&
         + 32/9\,\hat L^r_{5}\chi_{1}^2\,\chi_{6}\,{R^z_{14\eta}} \,{R^z_{\eta 61}}  - 32/9\,\hat L^r_{5}\chi_{1}^2\,\chi_{4}\,{R^z_{16\eta}} \,{R^z_{\eta 61}}  - 32/
         3\,\hat L^r_{5}\chi_{1}^3\,{R^z_{\eta 61}} ^2 
         \nonumber\\&&
         - 64/9\,(\chi_{6} - \chi_{4})\,\hat L^r_{7}\chi_{1}\,\chi_{6}\,{R^z_{14\eta}} \,{R^z_{\eta 61}}  - 128/9\,(\chi_{6} - \chi_{4})\,\hat L^r_{7}         \,\chi_{1}\,\chi_{4}\,{R^z_{16\eta}} \,{R^z_{\eta 61}} 
         \nonumber\\&&
         + 64/3\,(\chi_{6} - \chi_{4})\,\hat L^r_{7}\chi_{1}^2\,{R^z_{\eta 61}}  - 64/9\,(\chi_{6} + 2\,\chi_{4})\,\hat L^r_{6}\chi_{1}\,
         \chi_{6}\,{R^z_{14\eta}} \,{R^z_{\eta 61}}  
         \nonumber\\&&
         + 64/9\,(\chi_{6} + 2\,\chi_{4})\,\hat L^r_{6}\chi_{1}\,\chi_{4}\,{R^z_{16\eta}} \,{R^z_{\eta 61}}  + 64/3\,(\chi_{6} + 2\,\chi_{4})\,\hat L^r_{6}
         \chi_{1}^2\,{R^z_{\eta 61}} ^2
         \nonumber\\&&
         - 32/3\,(\chi_{6} + 2\,\chi_{4})\,\hat L^r_{4}\chi_{1}^2\,{R^z_{\eta 61}} ^2 - 32/9\,(\chi_{6} + 2\,\chi_{4})\,\hat L^r_{4}\chi_{1}^2\,
         {R^z_{16\eta}} \,{R^z_{\eta 61}} 
         \nonumber\\&&
         + 32/9\,(\chi_{6} + 2\,\chi_{4})\,\hat L^r_{4}\chi_{1}^2\,{R^z_{14\eta}} \,{R^z_{\eta 61}}  \Big)
       + {B}^{V}(\chi_\eta) \, \Big( 64/9\,(\chi_{6} - \chi_{4})^2\,\hat L^r_{7}\chi_{1}\,{R^z_{\eta 61}} ^2 
       \nonumber\\&&
       - 16/3\,(\chi_{6} + 2\,\chi_{4})\,\hat L^r_{4}\chi_{1}\,\chi_\eta\,
         {R^z_{\eta 61}} ^2 + 32/9\,(\chi_{6} + 2\,\chi_{4})\,(2\,\chi_{6} + \chi_{4})\,\hat L^r_{6}\chi_{1}\,{R^z_{\eta 61}} ^2 
         \nonumber\\&&
         - 16/9\,(2\,\chi_{6} + \chi_{4})\,\hat L^r_{5}
         \chi_{1}\,\chi_\eta\,{R^z_{\eta 61}} ^2 + 32/9\,(2\,\chi_{6}^2 + \chi_{4}^2)\,\hat L^r_{8}\chi_{1}\,{R^z_{\eta 61}} ^2 \Big)
       \nonumber\\&&
       + {C}^{V}(\chi_{1}) \, \Big(  - 32/3\,\hat L^r_{8}\chi_{1}^3\,{R^z_{146\eta}}  + 16/3\,\hat L^r_{5}\chi_{1}^3\,{R^z_{146\eta}}  - 32/3\,(\chi_{6} + 2\,\chi_{4})\,\hat L^r_{6}
         \chi_{1}^2\,{R^z_{146\eta}}  
         \nonumber\\&&
         + 16/3\,(\chi_{6} + 2\,\chi_{4})\,\hat L^r_{4}\chi_{1}^2\,{R^z_{146\eta}}  \Big)
       + {A}_{23}^{V}(\chi_{1}) \, \Big( 8\,\hat L^r_{3}\chi_{1}\,{R^c_{146\eta}}  - 12\,\hat L^r_{2}\chi_{1} - 24\,\hat L^r_{1}\chi_{1} + 8\,\hat L^r_{0}\chi_{1}\,{R^c_{146\eta}}  \Big)
       \nonumber\\&&
       + {A}_{23}^{V}(\chi_{14}) \, \Big(  - 24\,\hat L^r_{3}\chi_{1} - 48\,\hat L^r_{0}\chi_{1} \Big)
       + {A}_{23}^{V}(\chi_{16}) \, \Big(  - 12\,\hat L^r_{3}\chi_{1} - 24\,\hat L^r_{0}\chi_{1} \Big)
       \nonumber\\&&
       + {A}_{23}^{V}(\chi_{4}) \, \Big(  - 36\,\hat L^r_{2}\chi_{1} \Big)
       + {A}_{23}^{V}(\chi_{46}) \, \Big(  - 48\,\hat L^r_{2}\chi_{1} \Big)
       + {A}_{23}^{V}(\chi_\eta) \, \Big(  - 16\,\hat L^r_{3}\chi_{1}\,{R^z_{\eta 61}} ^2 - 12\,\hat L^r_{2}\chi_{1}
       \nonumber\\&&
       - 16\,\hat L^r_{0}\chi_{1}\,{R^z_{\eta 61}} ^2 \Big)
       + {B}_{23}^{V}(\chi_{1}) \, \Big( 8\,\hat L^r_{3}\chi_{1}\,{R^z_{146\eta}}  + 8\,\hat L^r_{0}\chi_{1}\,{R^z_{146\eta}}  \Big)
\end{eqnarray}
\begin{eqnarray}
\lefteqn{ F_0^4 \Delta^V\! m^{2(6R)12}_{12} =
       + \overline{A}(\chi_{1})\,{A}^{V}(\chi_{1}) \, \Big( \chi_{1} + 1/9\,\chi_{1}\,{R^c_{146\eta}} ^2 \Big)}&&
\nonumber\\&&
       + \overline{A}(\chi_{1})\,{A}^{V}(\chi_{14}) \, \Big( 4/45\,{R^z_{146\eta}}  + 8/9\,\chi_{1}\,{R^c_{146\eta}}  + 4/45\,(\chi_{4} - 11\,\chi_{1})\,{R^z_{16\eta}}  \Big)
\nonumber\\&&
       + \overline{A}(\chi_{1})\,{A}^{V}(\chi_{16}) \, \Big( 2/45\,{R^z_{146\eta}}  + 4/9\,\chi_{1}\,{R^c_{146\eta}}  + 2/45\,(\chi_{6} - 11\,\chi_{1})\,{R^z_{14\eta}}  \Big)
\nonumber\\&&
       + \overline{A}(\chi_{1})\,{A}^{V}(\chi_{46}) \, \Big(  - 2/27\,\chi_{1}\,{R^z_{16\eta}} ^2 + 4/27\,\chi_{1}\,{R^z_{14\eta}} \,{R^z_{16\eta}}  - 2/27\,\chi_{1}\,{R^z_{14\eta}} ^2 \Big)
\nonumber\\&&
       + \overline{A}(\chi_{1})\,{A}^{V}(\chi_\eta) \, \Big(  - 2/9\,\chi_{1}\,{R^c_{146\eta}} \,{R^z_{\eta 61}} ^2 \Big)
       + \overline{A}(\chi_{1})\,{B}^{V}(\chi_{1}) \, \Big( 1/9\,\chi_{1}\,{R^z_{146\eta}} \,{R^c_{146\eta}}  
       \nonumber\\&&
       + \chi_{1}^2 + 2/9\,\chi_{1}^2\,{R^c_{146\eta}} ^2 \Big)
       + \overline{A}(\chi_{1})\,{B}^{V}(\chi_{1},\chi_\eta) \, \Big(  - 8/9\,\chi_{1}^2\,{R^c_{146\eta}} \,{R^z_{\eta 61}} ^2 \Big)
       \nonumber\\&&
       + \overline{A}(\chi_{1})\,{C}^{V}(\chi_{1}) \, \Big( 2/9\,\chi_{1}^2\,{R^z_{146\eta}} \,{R^c_{146\eta}}  \Big)
       + \overline{A}(\chi_{14})\,{A}^{V}(\chi_{1}) \, \Big( 4/45\,{R^z_{146\eta}}  + 8/9\,\chi_{1}\,{R^c_{146\eta}} 
       \nonumber\\&&
       + 4/45\,(\chi_{4} - 11\,\chi_{1})\,{R^z_{16\eta}}  \Big)
       + \overline{A}(\chi_{14})\,{A}^{V}(\chi_{14}) \, \Big(  - 4/27\,{R^z_{146\eta}}  + 10/27\,\chi_{1}\,{R^z_{16\eta}}  
       \nonumber\\&&
       + 1/27\,(2\,\chi_\eta - 2\,\chi_{4} - 47\,\chi_{1}) +
         2/27\,(4\,\chi_\eta + 4\,\chi_{4} + 9\,\chi_{1})\,{R^z_{\eta 61}} ^2 + 2/27\,(4\,\chi_\eta + \chi_{1})\,{R^z_{\eta 61}}  
         \nonumber\\&&
         - 1/27\,(4\,\chi_{4} + 13\,
         \chi_{1})\,{R^c_{146\eta}}  \Big)
       + \overline{A}(\chi_{14})\,{A}^{V}(\chi_{16}) \, \Big(  - \chi_{1} \Big)
       + \overline{A}(\chi_{14})\,{A}^{V}(\chi_\eta) \, \Big(  - 2/45\,(\chi_\eta - \chi_{4})
       \nonumber\\&&
       - 8/45\,(\chi_\eta + 9\,\chi_{1})\,{R^z_{\eta 61}} ^2 - 4/45\,(2\,\chi_\eta
          - \chi_{4} + 9\,\chi_{1})\,{R^z_{\eta 61}}  \Big)
       + \overline{A}(\chi_{14})\,{B}^{V}(\chi_{1}) \, \Big( 8/9\,\chi_{1}\,{R^z_{146\eta}}  
       \nonumber\\&&
       - 4/9\,(\chi_{4} + 2\,\chi_{1})\,\chi_{1}\,{R^z_{16\eta}}  \Big)
       + \overline{A}(\chi_{14})\,{B}^{V}(\chi_{1},\chi_\eta) \, \Big(  - 4/9\,(\chi_{4} + 2\,\chi_{1})\,\chi_{1}\,{R^z_{\eta 61}}  \Big)
       \nonumber\\&&
       + \overline{A}(\chi_{16})\,{A}^{V}(\chi_{1}) \, \Big( 2/45\,{R^z_{146\eta}}  + 4/9\,\chi_{1}\,{R^c_{146\eta}}  + 2/45\,(\chi_{6} - 11\,\chi_{1})\,{R^z_{14\eta}}  \Big)
       \nonumber\\&&
       + \overline{A}(\chi_{16})\,{A}^{V}(\chi_{14}) \, \Big(  - \chi_{1} \Big)
       + \overline{A}(\chi_{16})\,{A}^{V}(\chi_{16}) \, \Big(  - 2/27\,{R^z_{146\eta}}  + 5/27\,\chi_{1}\,{R^z_{14\eta}}  
       \nonumber\\&&
       + 2/27\,(2\,\chi_\eta - 2\,\chi_{6} - 5\,\chi_{1}) + 1/
         27\,(4\,\chi_\eta + 4\,\chi_{6} + 9\,\chi_{1})\,{R^z_{\eta 61}} ^2 - 2/27\,(4\,\chi_\eta + \chi_{1})\,{R^z_{\eta 61}}  
         \nonumber\\&&
         - 1/54\,(4\,\chi_{6} + 13\,\chi_{1}
         \Big) \,{R^c_{146\eta}}  \Big)
       + \overline{A}(\chi_{16})\,{A}^{V}(\chi_\eta) \, \Big(  - 4/45\,(\chi_\eta - \chi_{6})
       \nonumber\\&&
       - 4/45\,(\chi_\eta + 9\,\chi_{1})\,{R^z_{\eta 61}} ^2 + 4/45\,(2\,\chi_\eta
          - \chi_{6} + 9\,\chi_{1})\,{R^z_{\eta 61}}  \Big)
       + \overline{A}(\chi_{16})\,{B}^{V}(\chi_{1}) \, \Big( 4/9\,\chi_{1}\,{R^z_{146\eta}} 
       \nonumber\\&&
       - 2/9\,(\chi_{6} + 2\,\chi_{1})\,\chi_{1}\,{R^z_{14\eta}}  \Big)
       + \overline{A}(\chi_{16})\,{B}^{V}(\chi_{1},\chi_\eta) \, \Big( 4/9\,(\chi_{6} + 2\,\chi_{1})\,\chi_{1}\,{R^z_{\eta 61}}  \Big)
       \nonumber\\&&
       + \overline{A}(\chi_{4})\,{B}^{V}(\chi_{1}) \, \Big( 1/3\,\chi_{1}\,\chi_{4}\,{R^z_{16\eta}} ^2 \Big)
       + \overline{A}(\chi_{4})\,{B}^{V}(\chi_{1},\chi_\eta) \, \Big( 2/3\,\chi_{1}\,\chi_{4}\,{R^z_{16\eta}} \,{R^z_{\eta 61}}  \Big)
       \nonumber\\&&
       + \overline{A}(\chi_{4})\,{B}^{V}(\chi_\eta) \, \Big( 1/3\,\chi_{1}\,\chi_{4}\,{R^z_{\eta 61}} ^2 \Big)
       + \overline{A}(\chi_{46})\,{A}^{V}(\chi_{1}) \, \Big(  - 2/27\,\chi_{1}\,{R^z_{16\eta}} ^2 + 4/27\,\chi_{1}\,{R^z_{14\eta}} \,{R^z_{16\eta}} 
       \nonumber\\&&
       - 2/27\,\chi_{1}\,{R^z_{14\eta}} ^2 \Big)
       + \overline{A}(\chi_{46})\,{A}^{V}(\chi_\eta) \, \Big(  - 2/3\,\chi_{1}\,{R^z_{\eta 61}} ^2 - 4/9\,\chi_{1}\,{R^z_{16\eta}} \,{R^z_{\eta 61}}  + 4/9\,\chi_{1}\,{R^z_{14\eta}} \,{R^z_{\eta 61}}  \Big)
       \nonumber\\&&
       + \overline{A}(\chi_{46})\,{B}^{V}(\chi_{1}) \, \Big( 4/27\,(\chi_{6} + \chi_{4} + \chi_{1})\,\chi_{1}\,{R^z_{14\eta}} \,{R^z_{16\eta}}  + 2/27\,(\chi_{6} - \chi_{1})\,\chi_{1}\,
         {R^z_{14\eta}} ^2 
         \nonumber\\&&
         + 2/27\,(\chi_{4} - \chi_{1})\,\chi_{1}\,{R^z_{16\eta}} ^2 \Big)
       + \overline{A}(\chi_{46})\,{B}^{V}(\chi_{1},\chi_\eta) \, \Big(  - 4/27\,(\chi_{6} - \chi_{4} - 3\,\chi_{1})\,\chi_{1}\,{R^z_{14\eta}} \,{R^z_{\eta 61}} 
       \nonumber\\&&
       - 4/27\,(2\,\chi_{6}
          + \chi_{4} + 3\,\chi_{1})\,\chi_{1}\,{R^z_{16\eta}} \,{R^z_{\eta 61}}  \Big)
       + \overline{A}(\chi_{46})\,{B}^{V}(\chi_\eta) \, \Big(  - 2/9\,(3\,\chi_\eta + \chi_{4})\,\chi_{1}\,{R^z_{\eta 61}} ^2 \Big)
       \nonumber\\&&
       + \overline{A}(\chi_\eta)\,{A}^{V}(\chi_{1}) \, \Big(  - 2/9\,\chi_{1}\,{R^c_{146\eta}} \,{R^z_{\eta 61}} ^2 \Big)
       + \overline{A}(\chi_\eta)\,{A}^{V}(\chi_{14}) \, \Big(  - 2/45\,(\chi_\eta - \chi_{4}) 
       \nonumber\\&&
       - 8/45\,(\chi_\eta + 9\,\chi_{1})\,{R^z_{\eta 61}} ^2 - 4/45\,(2\,\chi_\eta
          - \chi_{4} + 9\,\chi_{1})\,{R^z_{\eta 61}}  \Big)
       + \overline{A}(\chi_\eta)\,{A}^{V}(\chi_{16}) \, \Big(  - 4/45\,(\chi_\eta - \chi_{6}) 
       \nonumber\\&&
       - 4/45\,(\chi_\eta + 9\,\chi_{1})\,{R^z_{\eta 61}} ^2 + 4/45\,(2\,\chi_\eta
          - \chi_{6} + 9\,\chi_{1})\,{R^z_{\eta 61}}  \Big)
       + \overline{A}(\chi_\eta)\,{A}^{V}(\chi_{46}) \, \Big(  - 2/3\,\chi_{1}\,{R^z_{\eta 61}} ^2 
       \nonumber\\&&
       - 4/9\,\chi_{1}\,{R^z_{16\eta}} \,{R^z_{\eta 61}}  + 4/9\,\chi_{1}\,{R^z_{14\eta}} \,{R^z_{\eta 61}}  \Big)
       + \overline{A}(\chi_\eta)\,{A}^{V}(\chi_\eta) \, \Big( 4/9\,\chi_{1}\,{R^z_{\eta 61}} ^4 \Big)
       \nonumber\\&&
       + \overline{A}(\chi_\eta)\,{B}^{V}(\chi_{1}) \, \Big(  - 2/9\,\chi_{1}\,{R^z_{146\eta}} \,{R^z_{\eta 61}} ^2 + 2/27\,\chi_{1}\,\chi_{6}\,{R^z_{14\eta}} ^2 + 1/27\,\chi_{1}\,\chi_{4}\,
         {R^z_{16\eta}} ^2 
         \nonumber\\&&
         - 4/9\,\chi_{1}^2\,{R^c_{146\eta}} \,{R^z_{\eta 61}} ^2 \Big)
       + \overline{A}(\chi_\eta)\,{B}^{V}(\chi_{1},\chi_\eta) \, \Big(  - 8/27\,\chi_{1}\,\chi_{6}\,{R^z_{14\eta}} \,{R^z_{\eta 61}}  + 2/27\,\chi_{1}\,\chi_{4}\,{R^z_{16\eta}} \,{R^z_{\eta 61}} 
       \nonumber\\&&
       + 8/9\,
         \chi_{1}^2\,{R^z_{\eta 61}} ^4 + 4/9\,\chi_{1}^2\,{R^c_{146\eta}} \,{R^z_{\eta 61}} ^2 \Big)
       + \overline{A}(\chi_\eta)\,{B}^{V}(\chi_\eta) \, \Big( 1/27\,(8\,\chi_{6} + \chi_{4})\,\chi_{1}\,{R^z_{\eta 61}} ^2 \Big)
              \nonumber\\&&
       + \overline{A}(\chi_\eta)\,{C}^{V}(\chi_{1}) \, \Big(  - 4/9\,\chi_{1}^2\,{R^z_{146\eta}} \,{R^z_{\eta 61}} ^2 \Big)
       + \overline{B}(\chi_{1})\,{A}^{V}(\chi_{1}) \, \Big( 1/9\,\chi_{1}\,{R^z_{146\eta}} \,{R^c_{146\eta}}  + \chi_{1}^2 
              \nonumber\\&&
              + 2/9\,\chi_{1}^2\,{R^c_{146\eta}} ^2 \Big)
       + \overline{B}(\chi_{1})\,{A}^{V}(\chi_{14}) \, \Big( 8/9\,\chi_{1}\,{R^z_{146\eta}}  - 4/9\,(\chi_{4} + 2\,\chi_{1})\,\chi_{1}\,{R^z_{16\eta}}  \Big)
              \nonumber\\&&
       + \overline{B}(\chi_{1})\,{A}^{V}(\chi_{16}) \, \Big( 4/9\,\chi_{1}\,{R^z_{146\eta}}  - 2/9\,(\chi_{6} + 2\,\chi_{1})\,\chi_{1}\,{R^z_{14\eta}}  \Big)
       + \overline{B}(\chi_{1})\,{A}^{V}(\chi_{4}) \, \Big( 1/3\,\chi_{1}\,\chi_{4}\,{R^z_{16\eta}} ^2 \Big)
              \nonumber\\&&
       + \overline{B}(\chi_{1})\,{A}^{V}(\chi_{46}) \, \Big( 4/27\,(\chi_{6} + \chi_{4} + \chi_{1})\,\chi_{1}\,{R^z_{14\eta}} \,{R^z_{16\eta}}  + 2/27\,(\chi_{6} - \chi_{1})\,\chi_{1}\,
         {R^z_{14\eta}} ^2 
                \nonumber\\&&
                + 2/27\,(\chi_{4} - \chi_{1})\,\chi_{1}\,{R^z_{16\eta}} ^2 \Big)
       + \overline{B}(\chi_{1})\,{A}^{V}(\chi_\eta) \, \Big(  - 2/9\,\chi_{1}\,{R^z_{146\eta}} \,{R^z_{\eta 61}} ^2 + 2/27\,\chi_{1}\,\chi_{6}\,{R^z_{14\eta}} ^2
              \nonumber\\&&
              + 1/27\,\chi_{1}\,\chi_{4}\,
         {R^z_{16\eta}} ^2 - 4/9\,\chi_{1}^2\,{R^c_{146\eta}} \,{R^z_{\eta 61}} ^2 \Big)
       + \overline{B}(\chi_{1})\,{B}^{V}(\chi_{1}) \, \Big( 1/9\,\chi_{1}\,{R^z_{146\eta}} ^2 + 4/9\,\chi_{1}^2\,{R^z_{146\eta}} \,{R^c_{146\eta}}  \Big)
              \nonumber\\&&
       + \overline{B}(\chi_{1})\,{B}^{V}(\chi_{1},\chi_\eta) \, \Big(  - 4/9\,\chi_{1}^2\,{R^z_{146\eta}} \,{R^z_{\eta 61}} ^2 \Big)
       + \overline{B}(\chi_{1})\,{C}^{V}(\chi_{1}) \, \Big( 2/9\,\chi_{1}^2\,{R^z_{146\eta}} ^2 \Big)
              \nonumber\\&&
       + \overline{B}(\chi_{1},\chi_\eta)\,{A}^{V}(\chi_{14}) \, \Big(  - 4/9\,(\chi_{4} + 2\,\chi_{1})\,\chi_{1}\,{R^z_{\eta 61}}  \Big)
       + \overline{B}(\chi_{1},\chi_\eta)\,{A}^{V}(\chi_{16}) \, \Big( 4/9\,(\chi_{6} 
              \nonumber\\&&
              + 2\,\chi_{1})\,\chi_{1}\,{R^z_{\eta 61}}  \Big)
       + \overline{B}(\chi_{1},\chi_\eta)\,{A}^{V}(\chi_{4}) \, \Big( 2/3\,\chi_{1}\,\chi_{4}\,{R^z_{16\eta}} \,{R^z_{\eta 61}}  \Big)
              \nonumber\\&&
       + \overline{B}(\chi_{1},\chi_\eta)\,{A}^{V}(\chi_{46}) \, \Big(  - 4/27\,(\chi_{6} - \chi_{4} - 3\,\chi_{1})\,\chi_{1}\,{R^z_{14\eta}} \,{R^z_{\eta 61}}  - 4/27\,(2\,\chi_{6}
          + \chi_{4} 
                 \nonumber\\&&
                 + 3\,\chi_{1})\,\chi_{1}\,{R^z_{16\eta}} \,{R^z_{\eta 61}}  \Big)
       + \overline{B}(\chi_{1},\chi_\eta)\,{A}^{V}(\chi_\eta) \, \Big(  - 8/27\,\chi_{1}\,\chi_{6}\,{R^z_{14\eta}} \,{R^z_{\eta 61}}  + 2/27\,\chi_{1}\,\chi_{4}\,{R^z_{16\eta}} \,{R^z_{\eta 61}}  
              \nonumber\\&&
              + 8/9\,
         \chi_{1}^2\,{R^z_{\eta 61}} ^4 - 4/9\,\chi_{1}^2\,{R^c_{146\eta}} \,{R^z_{\eta 61}} ^2 \Big)
       + \overline{B}(\chi_{1},\chi_\eta)\,{B}^{V}(\chi_{1}) \, \Big(  - 4/9\,\chi_{1}^2\,{R^z_{146\eta}} \,{R^z_{\eta 61}} ^2 \Big)
              \nonumber\\&&
       + \overline{B}(\chi_\eta)\,{A}^{V}(\chi_{4}) \, \Big( 1/3\,\chi_{1}\,\chi_{4}\,{R^z_{\eta 61}} ^2 \Big)
       + \overline{B}(\chi_\eta)\,{A}^{V}(\chi_{46}) \, \Big(  - 2/9\,(3\,\chi_\eta + \chi_{4})\,\chi_{1}\,{R^z_{\eta 61}} ^2 \Big)
              \nonumber\\&&
       + \overline{B}(\chi_\eta)\,{A}^{V}(\chi_\eta) \, \Big( 1/27\,(8\,\chi_{6} + \chi_{4})\,\chi_{1}\,{R^z_{\eta 61}} ^2 \Big)
       + \overline{C}(\chi_{1})\,{A}^{V}(\chi_{1}) \, \Big( 2/9\,\chi_{1}^2\,{R^z_{146\eta}} \,{R^c_{146\eta}}  \Big)
              \nonumber\\&&
       + \overline{C}(\chi_{1})\,{A}^{V}(\chi_\eta) \, \Big(  - 4/9\,\chi_{1}^2\,{R^z_{146\eta}} \,{R^z_{\eta 61}} ^2 \Big)
       + \overline{C}(\chi_{1})\,{B}^{V}(\chi_{1}) \, \Big( 2/9\,\chi_{1}^2\,{R^z_{146\eta}} ^2 \Big)
              \nonumber\\&&
       + {A}^{V}(\chi_{1})\,\frac{1}{16\pi^2} \, \Big(  - 1/4\,\chi_{1}^2 + 1/6\,\chi_{1}^2\,{R^z_{16\eta}}  + 1/12\,\chi_{1}^2\,{R^z_{14\eta}}  \Big)
       + {A}^{V}(\chi_{1})^2 \, \Big( 1/2\,\chi_{1} + 1/18\,\chi_{1}\,{R^c_{146\eta}} ^2 \Big)
              \nonumber\\&&
       + {A}^{V}(\chi_{1})\,{A}^{V}(\chi_{14}) \, \Big( 4/45\,{R^z_{146\eta}}  + 8/9\,\chi_{1}\,{R^c_{146\eta}}  + 4/45\,(\chi_{4} - 11\,\chi_{1})\,{R^z_{16\eta}}  \Big)
              \nonumber\\&&
       + {A}^{V}(\chi_{1})\,{A}^{V}(\chi_{16}) \, \Big( 2/45\,{R^z_{146\eta}}  + 4/9\,\chi_{1}\,{R^c_{146\eta}}  + 2/45\,(\chi_{6} - 11\,\chi_{1})\,{R^z_{14\eta}}  \Big)
              \nonumber\\&&
       + {A}^{V}(\chi_{1})\,{A}^{V}(\chi_{46}) \, \Big(  - 2/27\,\chi_{1}\,{R^z_{16\eta}} ^2 + 4/27\,\chi_{1}\,{R^z_{14\eta}} \,{R^z_{16\eta}}  - 2/27\,\chi_{1}\,{R^z_{14\eta}} ^2 \Big)
              \nonumber\\&&
       + {A}^{V}(\chi_{1})\,{A}^{V}(\chi_\eta) \, \Big(  - 2/9\,\chi_{1}\,{R^c_{146\eta}} \,{R^z_{\eta 61}} ^2 \Big)
       + {A}^{V}(\chi_{1})\,{B}^{V}(\chi_{1}) \, \Big( 1/9\,\chi_{1}\,{R^z_{146\eta}} \,{R^c_{146\eta}}  + \chi_{1}^2 
              \nonumber\\&&
              + 2/9\,\chi_{1}^2\,{R^c_{146\eta}} ^2 \Big)
       + {A}^{V}(\chi_{1})\,{B}^{V}(\chi_{1},\chi_\eta) \, \Big(  - 4/9\,\chi_{1}^2\,{R^c_{146\eta}} \,{R^z_{\eta 61}} ^2 \Big)
              \nonumber\\&&
       + {A}^{V}(\chi_{1})\,{C}^{V}(\chi_{1}) \, \Big( 2/9\,\chi_{1}^2\,{R^z_{146\eta}} \,{R^c_{146\eta}}  \Big)
       + {A}^{V}(\chi_{14})\,\frac{1}{16\pi^2} \, \Big(  - 1/6\,\chi_{1}\,{R^z_{146\eta}}  + 1/36\,(3\,\chi_\eta + 18\,\chi_{6} 
              \nonumber\\&&
              + 42\,\chi_{4} - \chi_{1})\,\chi_{1} + 1/18\,
         \Big(6\,\chi_\eta + 3\,\chi_{4} + \chi_{1})\,\chi_{1}\,{R^z_{\eta 61}}  + 1/18\,(6\,\chi_\eta + 3\,\chi_{4} + \chi_{1})\,\chi_{1}\,{R^z_{\eta 61}} ^2 
                \nonumber\\&&
                + 1/18\,(3\,
         \chi_{4} + 7\,\chi_{1})\,\chi_{1}\,{R^z_{16\eta}}  - 1/36\,(3\,\chi_{4} + 7\,\chi_{1})\,\chi_{1}\,{R^c_{146\eta}}  \Big)
       + {A}^{V}(\chi_{14})^2 \, \Big(  - 2/27\,{R^z_{146\eta}} 
              \nonumber\\&&
              + 5/27\,\chi_{1}\,{R^z_{16\eta}}  + 1/54\,(2\,\chi_\eta - 2\,\chi_{4} - 47\,\chi_{1}) + 1/27\,
         \Big(4\,\chi_\eta + 4\,\chi_{4} + 9\,\chi_{1})\,{R^z_{\eta 61}} ^2 
                \nonumber\\&&
                + 1/27\,(4\,\chi_\eta + \chi_{1})\,{R^z_{\eta 61}}  - 1/54\,(4\,\chi_{4} + 13\,\chi_{1})\,
         {R^c_{146\eta}}  \Big)
       + {A}^{V}(\chi_{14})\,{A}^{V}(\chi_{16}) \, \Big(  - \chi_{1} \Big)
              \nonumber\\&&
       + {A}^{V}(\chi_{14})\,{A}^{V}(\chi_\eta) \, \Big(  - 2/45\,(\chi_\eta - \chi_{4}) - 8/45\,(\chi_\eta + 9\,\chi_{1})\,{R^z_{\eta 61}} ^2 - 4/45\,(2\,\chi_\eta
          - \chi_{4} + 9\,\chi_{1})\,{R^z_{\eta 61}}  \Big)
                 \nonumber\\&&
       + {A}^{V}(\chi_{14})\,{B}^{V}(\chi_{1}) \, \Big( 8/9\,\chi_{1}\,{R^z_{146\eta}}  - 4/9\,(\chi_{4} + 2\,\chi_{1})\,\chi_{1}\,{R^z_{16\eta}}  \Big)
              \nonumber\\&&
       + {A}^{V}(\chi_{14})\,{B}^{V}(\chi_{1},\chi_\eta) \, \Big(  - 4/9\,(\chi_{4} + 2\,\chi_{1})\,\chi_{1}\,{R^z_{\eta 61}}  \Big)
       + {A}^{V}(\chi_{16})\,\frac{1}{16\pi^2} \, \Big(  - 1/12\,\chi_{1}\,{R^z_{146\eta}}  
              \nonumber\\&&
              - 1/18\,(6\,\chi_\eta + 3\,\chi_{6} + \chi_{1})\,\chi_{1}\,{R^z_{\eta 61}}  + 1/36\,(6\,
         \chi_\eta + 3\,\chi_{6} + \chi_{1})\,\chi_{1}\,{R^z_{\eta 61}} ^2 + 1/72\,(12\,\chi_\eta + 15\,\chi_{6}
                \nonumber\\&&
                + 36\,\chi_{4} - \chi_{1})\,\chi_{1} + 1/36\,(3\,
         \chi_{6} + 7\,\chi_{1})\,\chi_{1}\,{R^z_{14\eta}}  - 1/72\,(3\,\chi_{6} + 7\,\chi_{1})\,\chi_{1}\,{R^c_{146\eta}}  \Big)
                \nonumber\\&&
       + {A}^{V}(\chi_{16})^2 \, \Big(  - 1/27\,{R^z_{146\eta}}  + 5/54\,\chi_{1}\,{R^z_{14\eta}}  + 1/27\,(2\,\chi_\eta - 2\,\chi_{6} - 5\,\chi_{1}) + 1/54\,(
         4\,\chi_\eta + 4\,\chi_{6} 
                \nonumber\\&&
                + 9\,\chi_{1})\,{R^z_{\eta 61}} ^2 - 1/27\,(4\,\chi_\eta + \chi_{1})\,{R^z_{\eta 61}}  - 1/108\,(4\,\chi_{6} + 13\,\chi_{1})\,
         {R^c_{146\eta}}  \Big)
                \nonumber\\&&
       + {A}^{V}(\chi_{16})\,{A}^{V}(\chi_\eta) \, \Big(  - 4/45\,(\chi_\eta - \chi_{6}) - 4/45\,(\chi_\eta + 9\,\chi_{1})\,{R^z_{\eta 61}} ^2 + 4/45\,(2\,\chi_\eta
          - \chi_{6} + 9\,\chi_{1})\,{R^z_{\eta 61}}  \Big)
                 \nonumber\\&&
       + {A}^{V}(\chi_{16})\,{B}^{V}(\chi_{1}) \, \Big( 4/9\,\chi_{1}\,{R^z_{146\eta}}  - 2/9\,(\chi_{6} + 2\,\chi_{1})\,\chi_{1}\,{R^z_{14\eta}}  \Big)
       + {A}^{V}(\chi_{16})\,{B}^{V}(\chi_{1},\chi_\eta) \, \Big( 4/9\,(\chi_{6} 
              \nonumber\\&&
              + 2\,\chi_{1})\,\chi_{1}\,{R^z_{\eta 61}}  \Big)
       + {A}^{V}(\chi_{4})\,{B}^{V}(\chi_{1}) \, \Big( 1/3\,\chi_{1}\,\chi_{4}\,{R^z_{16\eta}} ^2 \Big)
              \nonumber\\&&
       + {A}^{V}(\chi_{4})\,{B}^{V}(\chi_{1},\chi_\eta) \, \Big( 2/3\,\chi_{1}\,\chi_{4}\,{R^z_{16\eta}} \,{R^z_{\eta 61}}  \Big)
       + {A}^{V}(\chi_{4})\,{B}^{V}(\chi_\eta) \, \Big( 1/3\,\chi_{1}\,\chi_{4}\,{R^z_{\eta 61}} ^2 \Big)
                     \nonumber\\&&
       + {A}^{V}(\chi_{46})\,{A}^{V}(\chi_\eta) \, \Big(  - 2/3\,\chi_{1}\,{R^z_{\eta 61}} ^2 - 4/9\,\chi_{1}\,{R^z_{16\eta}} \,{R^z_{\eta 61}}  + 4/9\,\chi_{1}\,{R^z_{14\eta}} \,{R^z_{\eta 61}}  \Big)
                     \nonumber\\&&
       + {A}^{V}(\chi_{46})\,{B}^{V}(\chi_{1}) \, \Big( 4/27\,(\chi_{6} + \chi_{4} + \chi_{1})\,\chi_{1}\,{R^z_{14\eta}} \,{R^z_{16\eta}}  + 2/27\,(\chi_{6} - \chi_{1})\,\chi_{1}\,
         {R^z_{14\eta}} ^2 
                       \nonumber\\&&
                       + 2/27\,(\chi_{4} - \chi_{1})\,\chi_{1}\,{R^z_{16\eta}} ^2 \Big)
       + {A}^{V}(\chi_{46})\,{B}^{V}(\chi_{1},\chi_\eta) \, \Big(  - 4/27\,(\chi_{6} - \chi_{4} - 3\,\chi_{1})\,\chi_{1}\,{R^z_{14\eta}} \,{R^z_{\eta 61}}  
                     \nonumber\\&&
                     - 4/27\,(2\,\chi_{6}
          + \chi_{4} + 3\,\chi_{1})\,\chi_{1}\,{R^z_{16\eta}} \,{R^z_{\eta 61}}  \Big)
       + {A}^{V}(\chi_{46})\,{B}^{V}(\chi_\eta) \, \Big(  - 2/9\,(3\,\chi_\eta + \chi_{4})\,\chi_{1}\,{R^z_{\eta 61}} ^2 \Big)
                     \nonumber\\&&
       + {A}^{V}(\chi_\eta)^2 \, \Big( 2/9\,\chi_{1}\,{R^z_{\eta 61}} ^4 \Big)
       + {A}^{V}(\chi_\eta)\,{B}^{V}(\chi_{1}) \, \Big(  - 2/9\,\chi_{1}\,{R^z_{146\eta}} \,{R^z_{\eta 61}} ^2 + 2/27\,\chi_{1}\,\chi_{6}\,{R^z_{14\eta}} ^2 
       \nonumber\\&&
       + 1/27\,\chi_{1}\,\chi_{4}\,
         {R^z_{16\eta}} ^2 - 4/9\,\chi_{1}^2\,{R^c_{146\eta}} \,{R^z_{\eta 61}} ^2 \Big)
       + {A}^{V}(\chi_\eta)\,{B}^{V}(\chi_{1},\chi_\eta) \, \Big(  - 8/27\,\chi_{1}\,\chi_{6}\,{R^z_{14\eta}} \,{R^z_{\eta 61}} 
       \nonumber\\&&
       + 2/27\,\chi_{1}\,\chi_{4}\,{R^z_{16\eta}} \,{R^z_{\eta 61}}  + 8/9\,
         \chi_{1}^2\,{R^z_{\eta 61}} ^4 \Big)
       + {A}^{V}(\chi_\eta)\,{B}^{V}(\chi_\eta) \, \Big( 1/27\,(8\,\chi_{6} + \chi_{4})\,\chi_{1}\,{R^z_{\eta 61}} ^2 \Big)
       \nonumber\\&&
       + {A}^{V}(\chi_\eta)\,{C}^{V}(\chi_{1}) \, \Big(  - 4/9\,\chi_{1}^2\,{R^z_{146\eta}} \,{R^z_{\eta 61}} ^2 \Big)
       + {B}^{V}(\chi_{1})^2 \, \Big( 1/18\,\chi_{1}\,{R^z_{146\eta}} ^2 + 2/9\,\chi_{1}^2\,{R^z_{146\eta}} \,{R^c_{146\eta}}  \Big)
       \nonumber\\&&
       + {B}^{V}(\chi_{1})\,{B}^{V}(\chi_{1},\chi_\eta) \, \Big(  - 4/9\,\chi_{1}^2\,{R^z_{146\eta}} \,{R^z_{\eta 61}} ^2 \Big)
       + {B}^{V}(\chi_{1})\,{C}^{V}(\chi_{1}) \, \Big( 2/9\,\chi_{1}^2\,{R^z_{146\eta}} ^2 \Big)
       \nonumber\\&&
       + {A}_{23}^{V}(\chi_{1})\,\frac{1}{16\pi^2} \, \Big( 3/4\,\chi_{1} - 1/2\,\chi_{1}\,{R^z_{16\eta}}  - 1/4\,\chi_{1}\,{R^z_{14\eta}}  \Big)
       + {A}_{23}^{V}(\chi_{14})\,\frac{1}{16\pi^2} \, \Big( 1/6\,\chi_{1} - 1/3\,\chi_{1}\,{R^z_{\eta 61}}  
       \nonumber\\&&
       - 1/3\,\chi_{1}\,{R^z_{\eta 61}} ^2 - 1/3\,\chi_{1}\,{R^z_{16\eta}}  + 1/6\,\chi_{1}\,
         {R^c_{146\eta}}  \Big)
       + {A}_{23}^{V}(\chi_{16})\,\frac{1}{16\pi^2} \, \Big( 1/12\,\chi_{1} + 1/3\,\chi_{1}\,{R^z_{\eta 61}}  
       \nonumber\\&&
       - 1/6\,\chi_{1}\,{R^z_{\eta 61}} ^2 - 1/6\,\chi_{1}\,{R^z_{14\eta}}  + 1/12\,\chi_{1}\,
         {R^c_{146\eta}}  \Big)
       + {H}^{V}(1,\chi_{1},\chi_{1},\chi_{1},\chi_{1}) \, \Big( 1/3\,\chi_{1}^2 + 2/9\,\chi_{1}^2\,{R^c_{146\eta}} ^2 \Big)
       \nonumber\\&&
       + {H}^{V}(1,\chi_{1},\chi_{1},\chi_\eta,\chi_{1}) \, \Big(  - 8/9\,\chi_{1}^2\,{R^c_{146\eta}} \,{R^z_{\eta 61}} ^2 \Big)
       + {H}^{V}(1,\chi_{1},\chi_{14},\chi_{14},\chi_{1}) \, \Big(  - 11/18\,\chi_{1}\,{R^z_{146\eta}}  
       \nonumber\\&&
       - 5/6\,\chi_{1}^2\,{R^c_{146\eta}}  - 1/9\,(4\,\chi_{4} - 7\,\chi_{1})\,
         \chi_{1}\,{R^z_{16\eta}}  \Big)
       + {H}^{V}(1,\chi_{1},\chi_{16},\chi_{16},\chi_{1}) \, \Big(  - 11/36\,\chi_{1}\,{R^z_{146\eta}} 
       \nonumber\\&&
       - 5/12\,\chi_{1}^2\,{R^c_{146\eta}}  - 1/18\,(4\,\chi_{6} - 7\,\chi_{1}
         \Big) \,\chi_{1}\,{R^z_{14\eta}}  \Big)
       + {H}^{V}(1,\chi_{1},\chi_\eta,\chi_\eta,\chi_{1}) \, \Big( 8/9\,\chi_{1}^2\,{R^z_{\eta 61}} ^4 \Big)
       \nonumber\\&&
       + {H}^{V}(1,\chi_{14},\chi_{14},\chi_\eta,\chi_{1}) \, \Big( 1/27\,(\chi_\eta - \chi_{4})\,(\chi_\eta - \chi_{4} - 6\,\chi_{1}) + 4/27\,(\chi_\eta + 2\,\chi_{1})
         \,(\chi_\eta - 4\,\chi_{1})\,{R^z_{\eta 61}} ^2 
         \nonumber\\&&
         + 4/27\,(\chi_\eta^2 - \chi_{4}\,\chi_\eta - 4\,\chi_{1}\,\chi_\eta + \chi_{1}\,\chi_{4} - 6\,\chi_{1}^2)\,{R^z_{\eta 61}}
          \Big)
       + {H}^{V}(1,\chi_{16},\chi_{16},\chi_\eta,\chi_{1}) \, \Big( 2/27\,(\chi_\eta \nonumber\\&&
       - \chi_{6})\,(\chi_\eta - \chi_{6} - 6\,\chi_{1}) + 2/27\,(\chi_\eta + 2\,\chi_{1})
         \,(\chi_\eta - 4\,\chi_{1})\,{R^z_{\eta 61}} ^2 - 4/27\,(\chi_\eta^2 - \chi_{6}\,\chi_\eta - 4\,\chi_{1}\,\chi_\eta 
         \nonumber\\&&
         + \chi_{1}\,\chi_{6} - 6\,\chi_{1}^2)\,{R^z_{\eta 61}}
          \Big)
       + {H}^{V}(1,\chi_{4},\chi_{14},\chi_{14},\chi_{1}) \, \Big( 3/4\,\chi_{1}\,\chi_{4} \Big)
       + {H}^{V}(1,\chi_{46},\chi_{14},\chi_{16},\chi_{1}) \, \Big( 1/2\,(\chi_{6} 
       \nonumber\\&&
       + \chi_{4})\,\chi_{1} \Big)
       + {H}^{V}(1,\chi_\eta,\chi_{14},\chi_{14},\chi_{1}) \, \Big( 1/12\,\chi_{1}\,\chi_\eta + 1/3\,\chi_{1}\,\chi_\eta\,{R^z_{\eta 61}}  + 1/3\,\chi_{1}\,\chi_\eta\,{R^z_{\eta 61}} ^2 \Big)
       \nonumber\\&&
       + {H}^{V}(1,\chi_\eta,\chi_{16},\chi_{16},\chi_{1}) \, \Big( 1/6\,\chi_{1}\,\chi_\eta - 1/3\,\chi_{1}\,\chi_\eta\,{R^z_{\eta 61}}  + 1/6\,\chi_{1}\,\chi_\eta\,{R^z_{\eta 61}} ^2 \Big)
       \nonumber\\&&
       + {H}^{V}(2,\chi_{1},\chi_{1},\chi_{1},\chi_{1}) \, \Big( 4/9\,\chi_{1}^2\,{R^z_{146\eta}} \,{R^c_{146\eta}}  \Big)
       + {H}^{V}(2,\chi_{1},\chi_{1},\chi_\eta,\chi_{1}) \, \Big(  - 8/9\,\chi_{1}^2\,{R^z_{146\eta}} \,{R^z_{\eta 61}} ^2 \Big)
       \nonumber\\&&
       + {H}^{V}(2,\chi_{1},\chi_{14},\chi_{14},\chi_{1}) \, \Big(  - 5/6\,\chi_{1}^2\,{R^z_{146\eta}}  \Big)
       + {H}^{V}(2,\chi_{1},\chi_{16},\chi_{16},\chi_{1}) \, \Big(  - 5/12\,\chi_{1}^2\,{R^z_{146\eta}}  \Big)
       \nonumber\\&&
       + {H}^{V}(5,\chi_{1},\chi_{1},\chi_{1},\chi_{1}) \, \Big( 2/9\,\chi_{1}^2\,{R^z_{146\eta}} ^2 \Big)
       + {H}_1^{V}(1,\chi_{1},\chi_{14},\chi_{14},\chi_{1}) \, \Big( 4/9\,\chi_{1}\,{R^z_{146\eta}}  + 4/3\,\chi_{1}^2\,{R^c_{146\eta}}  
       \nonumber\\&&
       + 4/9\,(\chi_{4} - 4\,\chi_{1})\,\chi_{1}\,
         {R^z_{16\eta}}  \Big)
       + {H}_1^{V}(1,\chi_{1},\chi_{16},\chi_{16},\chi_{1}) \, \Big( 2/9\,\chi_{1}\,{R^z_{146\eta}}  + 2/3\,\chi_{1}^2\,{R^c_{146\eta}} 
       \nonumber\\&&
       + 2/9\,(\chi_{6} - 4\,\chi_{1})\,\chi_{1}\,
         {R^z_{14\eta}}  \Big)
       + {H}_1^{V}(1,\chi_{14},\chi_{14},\chi_\eta,\chi_{1}) \, \Big( 4/9\,(\chi_\eta - \chi_{4})\,\chi_{1} + 16/9\,(\chi_\eta 
       \nonumber\\&&
       + 2\,\chi_{1})\,\chi_{1}\,{R^z_{\eta 61}} ^2 + 8/9
         \,(2\,\chi_\eta - \chi_{4} + 2\,\chi_{1})\,\chi_{1}\,{R^z_{\eta 61}}  \Big)
       + {H}_1^{V}(1,\chi_{16},\chi_{16},\chi_\eta,\chi_{1}) \, \Big( 8/9\,(\chi_\eta - \chi_{6})\,\chi_{1} 
                       \nonumber\\&&
       + 8/9\,(\chi_\eta + 2\,\chi_{1})\,\chi_{1}\,{R^z_{\eta 61}} ^2 - 8/9\,
         \Big(2\,\chi_\eta - \chi_{6} + 2\,\chi_{1})\,\chi_{1}\,{R^z_{\eta 61}}  \Big)
         \nonumber\\&&
       + {H}_1^{V}(2,\chi_{1},\chi_{14},\chi_{14},\chi_{1}) \, \Big( 4/3\,\chi_{1}^2\,{R^z_{146\eta}}  \Big)
       + {H}_1^{V}(2,\chi_{1},\chi_{16},\chi_{16},\chi_{1}) \, \Big( 2/3\,\chi_{1}^2\,{R^z_{146\eta}}  \Big)
                       \nonumber\\&&
       + {H}_{21}^{V}(1,\chi_{1},\chi_{14},\chi_{14},\chi_{1}) \, \Big( \chi_{1}^2\,{R^z_{16\eta}}  - 1/2\,\chi_{1}^2\,{R^c_{146\eta}}  \Big)
       + {H}_{21}^{V}(1,\chi_{1},\chi_{16},\chi_{16},\chi_{1}) \, \Big( 1/2\,\chi_{1}^2\,{R^z_{14\eta}} 
                       \nonumber\\&&
                       - 1/4\,\chi_{1}^2\,{R^c_{146\eta}}  \Big)
       + {H}_{21}^{V}(1,\chi_{4},\chi_{14},\chi_{14},\chi_{1}) \, \Big( 9/4\,\chi_{1}^2 \Big)
       + {H}_{21}^{V}(1,\chi_{46},\chi_{14},\chi_{16},\chi_{1}) \, \Big( 3\,\chi_{1}^2 \Big)
                       \nonumber\\&&
       + {H}_{21}^{V}(1,\chi_\eta,\chi_{14},\chi_{14},\chi_{1}) \, \Big( 1/4\,\chi_{1}^2 + \chi_{1}^2\,{R^z_{\eta 61}}  + \chi_{1}^2\,{R^z_{\eta 61}} ^2 \Big)
       + {H}_{21}^{V}(1,\chi_\eta,\chi_{16},\chi_{16},\chi_{1}) \, \Big( 1/2\,\chi_{1}^2 
                       \nonumber\\&&
                       - \chi_{1}^2\,{R^z_{\eta 61}}  + 1/2\,\chi_{1}^2\,{R^z_{\eta 61}} ^2 \Big)
       + {H}_{21}^{V}(2,\chi_{1},\chi_{14},\chi_{14},\chi_{1}) \, \Big(  - 1/2\,\chi_{1}^2\,{R^z_{146\eta}}  \Big)
                       \nonumber\\&&
       + {H}_{21}^{V}(2,\chi_{1},\chi_{16},\chi_{16},\chi_{1}) \, \Big(  - 1/4\,\chi_{1}^2\,{R^z_{146\eta}}  \Big)
       + {H}_{27}^{V}(1,\chi_{1},\chi_{14},\chi_{14},\chi_{1}) \, \Big(  - \chi_{1}\,{R^z_{16\eta}}  
                       \nonumber\\&&
                       + 1/2\,\chi_{1}\,{R^c_{146\eta}}  \Big)
       + {H}_{27}^{V}(1,\chi_{1},\chi_{16},\chi_{16},\chi_{1}) \, \Big(  - 1/2\,\chi_{1}\,{R^z_{14\eta}}  + 1/4\,\chi_{1}\,{R^c_{146\eta}}  \Big)
                       \nonumber\\&&
       + {H}_{27}^{V}(1,\chi_{4},\chi_{14},\chi_{14},\chi_{1}) \, \Big(  - 9/4\,\chi_{1} \Big)
       + {H}_{27}^{V}(1,\chi_{46},\chi_{14},\chi_{16},\chi_{1}) \, \Big(  - 3\,\chi_{1} \Big)
                       \nonumber\\&&
       + {H}_{27}^{V}(1,\chi_\eta,\chi_{14},\chi_{14},\chi_{1}) \, \Big(  - 1/4\,\chi_{1} - \chi_{1}\,{R^z_{\eta 61}}  - \chi_{1}\,{R^z_{\eta 61}} ^2 \Big)
                              \nonumber\\&&
       + {H}_{27}^{V}(1,\chi_\eta,\chi_{16},\chi_{16},\chi_{1}) \, \Big(  - 1/2\,\chi_{1} 
                       + \chi_{1}\,{R^z_{\eta 61}}  - 1/2\,\chi_{1}\,{R^z_{\eta 61}} ^2 \Big)
                                              \nonumber\\&&
       + {H}_{27}^{V}(2,\chi_{1},\chi_{14},\chi_{14},\chi_{1}) \, \Big( 1/2\,\chi_{1}\,{R^z_{146\eta}}  \Big)
       + {H}_{27}^{V}(2,\chi_{1},\chi_{16},\chi_{16},\chi_{1}) \, \Big( 1/4\,\chi_{1}\,{R^z_{146\eta}}  \Big)
\end{eqnarray}

\section{Expressions for the decay constant}
\label{appdecay}

\begin{equation}
F_0 \Delta^V\! F^{2(4)12}_{12} =
       + {A}^{V}(\chi_{14}) \, \Big( 1 \Big)
       + {A}^{V}(\chi_{16}) \, \Big( 1/2 \Big) 
\end{equation}
\begin{eqnarray}
\lefteqn{
   F_0^3 \Delta^V\! F^{2(6L)12}_{12} =
       + {A}^{V}(\chi_{1}) \, \Big(  - 4/3\,\hat L^r_{5}\,\chi_{1}\,{R^c_{146\eta}}  + 4\,\hat L^r_{3}\,{R^z_{146\eta}}  + 4\,\hat L^r_{3}\,\chi_{1}\,{R^c_{146\eta}}  - 10\,\hat L^r_{2}\,\chi_{1} - 4\,\hat L^r_{1}\,\chi_{1}
}&&
\nonumber\\&&
          + 4\,\hat L^r_{0}\,{R^z_{146\eta}}
          + 4\,\hat L^r_{0}\,\chi_{1}\,{R^c_{146\eta}}  \Big)
       + {A}^{V}(\chi_{14}) \, \Big( 4\,\hat L^r_{5}\,\chi_{1} - 4\,(\chi_{6} + 2\,\chi_{4})\,\hat L^r_{4} - 10\,(\chi_{4} + \chi_{1})\,\hat L^r_{3} 
\nonumber\\&&
       - 4\,(\chi_{4} + \chi_{1})\,\hat L^r_{0} \Big) + {A}^{V}(\chi_{16}) \, \Big( 2\,\hat L^r_{5}\,\chi_{1} - 2\,(\chi_{6} + 2\,\chi_{4})\,\hat L^r_{4} - 5\,(\chi_{6} + \chi_{1})\,\hat L^r_{3} - 2\,(\chi_{6} + \chi_{1})\,\hat L^r_{0} \Big)
\nonumber\\&&
       + {A}^{V}(\chi_{4}) \, \Big( 12\,\hat L^r_{4}\,\chi_{4} - 6\,\hat L^r_{2}\,\chi_{4} - 24\,\hat L^r_{1}\,\chi_{4} \Big)
       + {A}^{V}(\chi_{46}) \, \Big( 8\,(\chi_{6} + \chi_{4})\,\hat L^r_{4} - 4\,(\chi_{6} + \chi_{4})\,\hat L^r_{2} 
\nonumber\\&&
       - 16\,(\chi_{6} + \chi_{4})\,\hat L^r_{1} \Big) + {A}^{V}(\chi_\eta) \, \Big( 8/3\,\hat L^r_{5}\,\chi_{1}\,{R^z_{\eta 61}} ^2 - 8\,\hat L^r_{3}\,\chi_\eta\,{R^z_{\eta 61}} ^2 - 2\,\hat L^r_{2}\,\chi_\eta - 8\,\hat L^r_{1}\,\chi_\eta
\nonumber\\&&
 - 8\,\hat L^r_{0}\,\chi_\eta\,{R^z_{\eta 61}} ^2 + 4/3\,(2\,
         \chi_{6} + \chi_{4})\,\hat L^r_{4} \Big)
       + {B}^{V}(\chi_{1}) \, \Big(  - 4/3\,\hat L^r_{5}\,\chi_{1}\,{R^z_{146\eta}}  + 4\,\hat L^r_{3}\,\chi_{1}\,{R^z_{146\eta}}  
\nonumber\\&&
       + 4\,\hat L^r_{0}\,\chi_{1}\,{R^z_{146\eta}}  \Big) + {B}^{V}(\chi_{14}) \, \Big( 8\,(\chi_{4} + \chi_{1})\,(\chi_{6} + 2\,\chi_{4})\,\hat L^r_{6} - 4\,(\chi_{4} + \chi_{1})\,(\chi_{6} + 2\,\chi_{4})\,\hat L^r_{4} 
\nonumber\\&&
+ 4\,(\chi_{4} + \chi_{1})^2\,\hat L^r_{8}
          - 2\,(\chi_{4} + \chi_{1})^2\,\hat L^r_{5} \Big)
       + {B}^{V}(\chi_{16}) \, \Big( 4\,(\chi_{6} + \chi_{1})\,(\chi_{6} + 2\,\chi_{4})\,\hat L^r_{6} 
\nonumber\\&&
  - 2\,(\chi_{6} + \chi_{1})\,(\chi_{6} + 2\,\chi_{4})\,\hat L^r_{4} + 2\,(\chi_{6} + \chi_{1})^2\,\hat L^r_{8}
          - \Big(\chi_{6} + \chi_{1})^2\,\hat L^r_{5} \Big)
       + {A}_{23}^{V}(\chi_{1}) \, \Big(  - 4\,\hat L^r_{3}\,{R^c_{146\eta}}  
\nonumber\\&&
+ 6\,\hat L^r_{2} + 12\,\hat L^r_{1} - 4\,\hat L^r_{0}\,{R^c_{146\eta}}  \Big)
       + {A}_{23}^{V}(\chi_{14}) \, \Big( 12\,\hat L^r_{3} + 24\,\hat L^r_{0} \Big)
       + {A}_{23}^{V}(\chi_{16}) \, \Big( 6\,\hat L^r_{3} + 12\,\hat L^r_{0} \Big)
\nonumber\\&&
       + {A}_{23}^{V}(\chi_{4}) \, \Big( 18\,\hat L^r_{2} \Big)
       + {A}_{23}^{V}(\chi_{46}) \, \Big( 24\,\hat L^r_{2} \Big)
       + {A}_{23}^{V}(\chi_\eta) \, \Big( 8\,\hat L^r_{3}\,{R^z_{\eta 61}} ^2 + 6\,\hat L^r_{2} + 8\,\hat L^r_{0}\,{R^z_{\eta 61}} ^2 \Big)
\nonumber\\&&
       + {B}_{23}^{V}(\chi_{1}) \, \Big(  - 4\,\hat L^r_{3}\,{R^z_{146\eta}}  - 4\,\hat L^r_{0}\,{R^z_{146\eta}}  \Big)
\end{eqnarray}
\begin{eqnarray}
\lefteqn{
   F_0^3 \Delta^V\! F^{2(6R)12}_{12} =
       + \overline{A}(\chi_{1})\,{B}^{V}(\chi_{14}) \, \Big( 1/18\,{R^z_{146\eta}}  - 1/9\,(\chi_{4} + 2\,\chi_{1})\,{R^z_{16\eta}}  \Big)
}&&
\nonumber\\&&
       + \overline{A}(\chi_{1})\,{B}^{V}(\chi_{16}) \, \Big( 1/36\,{R^z_{146\eta}}  - 1/18\,(\chi_{6} + 2\,\chi_{1})\,{R^z_{14\eta}}  \Big)
\nonumber\\&&
       + \overline{A}(\chi_{14})\,{A}^{V}(\chi_{14}) \, \Big( 5/54 - 5/27\,{R^z_{\eta 61}}  - 5/27\,{R^z_{\eta 61}} ^2 - 5/27\,{R^z_{16\eta}}  + 5/54\,{R^c_{146\eta}}  \Big)
\nonumber\\&&
       + \overline{A}(\chi_{16})\,{A}^{V}(\chi_{16}) \, \Big( 5/108 + 5/27\,{R^z_{\eta 61}}  - 5/54\,{R^z_{\eta 61}} ^2 - 5/54\,{R^z_{14\eta}}  + 5/108\,{R^c_{146\eta}}  \Big)
\nonumber\\&&
       + \overline{A}(\chi_\eta)\,{B}^{V}(\chi_{14}) \, \Big(  - 1/36\,(\chi_\eta - \chi_{4}) - 1/9\,(\chi_\eta + \chi_{4} + \chi_{1})\,{R^z_{\eta 61}}  - 1/9\,(\chi_\eta - \chi_{1})\,{R^z_{\eta 61}} ^2 \Big)
\nonumber\\&&
       + \overline{A}(\chi_\eta)\,{B}^{V}(\chi_{16}) \, \Big(  - 1/18\,(\chi_\eta - \chi_{6}) + 1/9\,(\chi_\eta + \chi_{6} + \chi_{1})\,{R^z_{\eta 61}}  - 1/18\,(\chi_\eta - \chi_{1})\,{R^z_{\eta 61}} ^2 \Big)
\nonumber\\&&
       + \overline{B}(\chi_{14})\,{A}^{V}(\chi_{1}) \, \Big( 1/18\,{R^z_{146\eta}}  - 1/9\,(\chi_{4} + 2\,\chi_{1})\,{R^z_{16\eta}}  \Big)
       + \overline{B}(\chi_{14})\,{A}^{V}(\chi_\eta) \, \Big(  - 1/36\,(\chi_\eta - \chi_{4})
\nonumber\\&&
 - 1/9\,(\chi_\eta + \chi_{4} + \chi_{1})\,{R^z_{\eta 61}}  - 1/9\,(\chi_\eta - \chi_{1})\,{R^z_{\eta 61}} ^2 \Big)
       + \overline{B}(\chi_{16})\,{A}^{V}(\chi_{1}) \, \Big( 1/36\,{R^z_{146\eta}} 
\nonumber\\&&
 - 1/18\,(\chi_{6} + 2\,\chi_{1})\,{R^z_{14\eta}}  \Big)
       + \overline{B}(\chi_{16})\,{A}^{V}(\chi_\eta) \, \Big(  - 1/18\,(\chi_\eta - \chi_{6}) + 1/9\,(\chi_\eta + \chi_{6} + \chi_{1})\,{R^z_{\eta 61}} 
\nonumber\\&&
    - 1/18\,(\chi_\eta - \chi_{1})\,{R^z_{\eta 61}} ^2 \Big)
    + {A}^{V}(\chi_{1})\,\frac{1}{16\pi^2} \, \Big( 1/8\,\chi_{1} - 1/12\,\chi_{1}\,{R^z_{16\eta}}  - 1/24\,\chi_{1}\,{R^z_{14\eta}}  \Big)
\nonumber\\&&
       + {A}^{V}(\chi_{1})\,{B}^{V}(\chi_{14}) \, \Big( 1/18\,{R^z_{146\eta}}  - 1/9\,(\chi_{4} + 2\,\chi_{1})\,{R^z_{16\eta}}  \Big)
       + {A}^{V}(\chi_{1})\,{B}^{V}(\chi_{16}) \, \Big( 1/36\,{R^z_{146\eta}} 
\nonumber\\&&
 - 1/18\,(\chi_{6} + 2\,\chi_{1})\,{R^z_{14\eta}}  \Big)
       + {A}^{V}(\chi_{14})\,\frac{1}{16\pi^2} \, \Big( 1/12\,{R^z_{146\eta}}  - 1/24\,(\chi_\eta + 6\,\chi_{6} + 14\,\chi_{4} + 5\,\chi_{1}) 
\nonumber\\&&
         - 1/12\,(2\,\chi_\eta + \chi_{4} + \chi_{1})\,{R^z_{\eta 61}}  - 1/12\,(2\,\chi_\eta + \chi_{4} + \chi_{1})\,{R^z_{\eta 61}} ^2 - 1/12\,(\chi_{4} + 3\,\chi_{1})\,{R^z_{16\eta}}  
\nonumber\\&&
       + 1/24\,(\chi_{4} + 3\,\chi_{1})\,{R^c_{146\eta}}  \Big) + {A}^{V}(\chi_{14})^2 \, \Big( 5/108 - 5/54\,{R^z_{\eta 61}}  - 5/54\,{R^z_{\eta 61}} ^2 - 5/54\,{R^z_{16\eta}}  
\nonumber\\&&
       + 5/108\,{R^c_{146\eta}}  \Big) + {A}^{V}(\chi_{16})\,\frac{1}{16\pi^2} \, \Big( 1/24\,{R^z_{146\eta}}  + 1/12\,(2\,\chi_\eta + \chi_{6} + \chi_{1})\,{R^z_{\eta 61}} 
\nonumber\\&&
 - 1/24\,(2\,\chi_\eta + \chi_{6} + \chi_{1})\,{R^z_{\eta 61}} ^2 - 1/48\,
         \Big(4\,\chi_\eta + 5\,\chi_{6} + 12\,\chi_{4} + 5\,\chi_{1}) - 1/24\,(\chi_{6} + 3\,\chi_{1})\,{R^z_{14\eta}} 
\nonumber\\&&
        + 1/48\,(\chi_{6} + 3\,\chi_{1})\,{R^c_{146\eta}}  \Big) + {A}^{V}(\chi_{16})^2 \, \Big( 5/216 + 5/54\,{R^z_{\eta 61}}  - 5/108\,{R^z_{\eta 61}} ^2 - 5/108\,{R^z_{14\eta}}  
\nonumber\\&&
       + 5/216\,{R^c_{146\eta}}  \Big) + {A}^{V}(\chi_\eta)\,{B}^{V}(\chi_{14}) \, \Big(  - 1/36\,(\chi_\eta - \chi_{4}) - 1/9\,(\chi_\eta + \chi_{4} + \chi_{1})\,{R^z_{\eta 61}}  
\nonumber\\&&
       - 1/9\,(\chi_\eta - \chi_{1})\,{R^z_{\eta 61}} ^2 \Big) + {A}^{V}(\chi_\eta)\,{B}^{V}(\chi_{16}) \, \Big(  - 1/18\,(\chi_\eta - \chi_{6}) + 1/9\,(\chi_\eta + \chi_{6} + \chi_{1})\,{R^z_{\eta 61}} 
\nonumber\\&&
 - 1/18\,(\chi_\eta - \chi_{1})\,{R^z_{\eta 61}} ^2 \Big)
       + {A}_{23}^{V}(\chi_{1})\,\frac{1}{16\pi^2} \, \Big(  - 3/8 + 1/4\,{R^z_{16\eta}}  + 1/8\,{R^z_{14\eta}}  \Big)
\nonumber\\&&
       + {A}_{23}^{V}(\chi_{14})\,\frac{1}{16\pi^2} \, \Big(  - 1/12 + 1/6\,{R^z_{\eta 61}}  + 1/6\,{R^z_{\eta 61}} ^2 + 1/6\,{R^z_{16\eta}}  - 1/12\,{R^c_{146\eta}}  \Big)
\nonumber\\&&
       + {A}_{23}^{V}(\chi_{16})\,\frac{1}{16\pi^2} \, \Big(  - 1/24 - 1/6\,{R^z_{\eta 61}}  + 1/12\,{R^z_{\eta 61}} ^2 + 1/12\,{R^z_{14\eta}}  - 1/24\,{R^c_{146\eta}}  \Big)
\nonumber\\&&
       + {H}^{V}(1,\chi_{1},\chi_{14},\chi_{14},\chi_{1}) \, \Big( 1/12\,{R^z_{146\eta}}  - 1/6\,\chi_{1}\,{R^z_{16\eta}}  + 1/12\,\chi_{1}\,{R^c_{146\eta}}  \Big)
\nonumber\\&&
       + {H}^{V}(1,\chi_{1},\chi_{16},\chi_{16},\chi_{1}) \, \Big( 1/24\,{R^z_{146\eta}}  - 1/12\,\chi_{1}\,{R^z_{14\eta}}  + 1/24\,\chi_{1}\,{R^c_{146\eta}}  \Big)
\nonumber\\&&
       + {H}^{V}(1,\chi_{4},\chi_{14},\chi_{14},\chi_{1}) \, \Big(  - 3/8\,\chi_{4} \Big)
       + {H}^{V}(1,\chi_{46},\chi_{14},\chi_{16},\chi_{1}) \, \Big(  - 1/4\,(\chi_{6} + \chi_{4}) \Big)
\nonumber\\&&
       + {H}^{V}(1,\chi_\eta,\chi_{14},\chi_{14},\chi_{1}) \, \Big(  - 1/24\,\chi_\eta - 1/6\,\chi_\eta\,{R^z_{\eta 61}}  - 1/6\,\chi_\eta\,{R^z_{\eta 61}} ^2 \Big)
\nonumber\\&&
       + {H}^{V}(1,\chi_\eta,\chi_{16},\chi_{16},\chi_{1}) \, \Big(  - 1/12\,\chi_\eta + 1/6\,\chi_\eta\,{R^z_{\eta 61}}  - 1/12\,\chi_\eta\,{R^z_{\eta 61}} ^2 \Big)
\nonumber\\&&
       + {H}^{V}(2,\chi_{1},\chi_{14},\chi_{14},\chi_{1}) \, \Big( 5/54\,\chi_{1}\,{R^z_{146\eta}}  \Big)
       + {H}^{V}(2,\chi_{1},\chi_{16},\chi_{16},\chi_{1}) \, \Big( 5/108\,\chi_{1}\,{R^z_{146\eta}}  \Big)
\nonumber\\&&
       + {H}_1^{V}(2,\chi_{1},\chi_{14},\chi_{14},\chi_{1}) \, \Big(  - 1/108\,\chi_{1}\,{R^z_{146\eta}}  \Big)
       + {H}_1^{V}(2,\chi_{1},\chi_{16},\chi_{16},\chi_{1}) \, \Big(  - 1/216\,\chi_{1}\,{R^z_{146\eta}}  \Big)
\nonumber\\&&
       + {H}_1^{V}(3,\chi_{14},\chi_{1},\chi_{14},\chi_{1}) \, \Big(  - 1/54\,\chi_{1}\,{R^z_{146\eta}}  \Big)
       + {H}_1^{V}(3,\chi_{16},\chi_{1},\chi_{16},\chi_{1}) \, \Big(  - 1/108\,\chi_{1}\,{R^z_{146\eta}}  \Big)
\nonumber\\&&
       + {H}_{27}^{V}(1,\chi_{1},\chi_{14},\chi_{14},\chi_{1}) \, \Big( 1/2\,{R^z_{16\eta}}  - 1/4\,{R^c_{146\eta}}  \Big)
       + {H}_{27}^{V}(1,\chi_{1},\chi_{16},\chi_{16},\chi_{1}) \, \Big( 1/4\,{R^z_{14\eta}}  - 1/8\,{R^c_{146\eta}}  \Big)
\nonumber\\&&
       + {H}_{27}^{V}(1,\chi_{4},\chi_{14},\chi_{14},\chi_{1}) \, \Big( 9/8 \Big)
       + {H}_{27}^{V}(1,\chi_{46},\chi_{14},\chi_{16},\chi_{1}) \, \Big( 3/2 \Big)
\nonumber\\&&
       + {H}_{27}^{V}(1,\chi_\eta,\chi_{14},\chi_{14},\chi_{1}) \, \Big( 1/8 + 1/2\,{R^z_{\eta 61}}  + 1/2\,{R^z_{\eta 61}} ^2 \Big)
\nonumber\\&&
       + {H}_{27}^{V}(1,\chi_\eta,\chi_{16},\chi_{16},\chi_{1}) \, \Big( 1/4 - 1/2\,{R^z_{\eta 61}}  + 1/4\,{R^z_{\eta 61}} ^2 \Big)
\nonumber\\&&
       + {H}_{27}^{V}(2,\chi_{1},\chi_{14},\chi_{14},\chi_{1}) \, \Big(  - 1/4\,{R^z_{146\eta}}  \Big)
       + {H}_{27}^{V}(2,\chi_{1},\chi_{16},\chi_{16},\chi_{1}) \, \Big(  - 1/8\,{R^z_{146\eta}}  \Big)
\nonumber\\&&
       + {H}^{'V}(1,\chi_{1},\chi_{1},\chi_{1},\chi_{1}) \, \Big( 1/6\,\chi_{1}^2 + 1/9\,\chi_{1}^2\,{R^c_{146\eta}} ^2 \Big)
       + {H}^{'V}(1,\chi_{1},\chi_{1},\chi_\eta,\chi_{1}) \, \Big(  - 4/9\,\chi_{1}^2\,{R^c_{146\eta}} \,{R^z_{\eta 61}} ^2 \Big)
\nonumber\\&&
       + {H}^{'V}(1,\chi_{1},\chi_{14},\chi_{14},\chi_{1}) \, \Big(  - 11/36\,\chi_{1}\,{R^z_{146\eta}}  - 5/12\,\chi_{1}^2\,{R^c_{146\eta}}  - 1/18\,(4\,\chi_{4} - 7\,\chi_{1})\,\chi_{1}\,{R^z_{16\eta}}  \Big)
\nonumber\\&&
       + {H}^{'V}(1,\chi_{1},\chi_{16},\chi_{16},\chi_{1}) \, \Big(  - 11/72\,\chi_{1}\,{R^z_{146\eta}}  - 5/24\,\chi_{1}^2\,{R^c_{146\eta}}  - 1/36\,(4\,\chi_{6} - 7\,\chi_{1})\,\chi_{1}\,{R^z_{14\eta}}  \Big)
\nonumber\\&&
       + {H}^{'V}(1,\chi_{1},\chi_\eta,\chi_\eta,\chi_{1}) \, \Big( 4/9\,\chi_{1}^2\,{R^z_{\eta 61}} ^4 \Big)
       + {H}^{'V}(1,\chi_{14},\chi_{14},\chi_\eta,\chi_{1}) \, \Big( 1/54\,(\chi_\eta - \chi_{4})\,(\chi_\eta - \chi_{4} 
\nonumber\\&&
- 6\,\chi_{1}) + 2/27\,(\chi_\eta + 2\,\chi_{1})\,(\chi_\eta - 4\,\chi_{1})\,
         {R^z_{\eta 61}} ^2 + 2/27\,(\chi_\eta^2 - \chi_{4}\,\chi_\eta - 4\,\chi_{1}\,\chi_\eta + \chi_{1}\,\chi_{4} - 6\,\chi_{1}^2)\,{R^z_{\eta 61}}  \Big)
\nonumber\\&&
       + {H}^{'V}(1,\chi_{16},\chi_{16},\chi_\eta,\chi_{1}) \, \Big( 1/27\,(\chi_\eta - \chi_{6})\,(\chi_\eta - \chi_{6} - 6\,\chi_{1}) + 1/27\,(\chi_\eta + 2\,\chi_{1})\,(\chi_\eta - 4\,\chi_{1})\,
\nonumber\\&&
         {R^z_{\eta 61}} ^2 - 2/27\,(\chi_\eta^2 - \chi_{6}\,\chi_\eta - 4\,\chi_{1}\,\chi_\eta + \chi_{1}\,\chi_{6} - 6\,\chi_{1}^2)\,{R^z_{\eta 61}}  \Big)
       + {H}^{'V}(1,\chi_{4},\chi_{14},\chi_{14},\chi_{1}) \, \Big( 3/8\,\chi_{1}\,\chi_{4} \Big)
\nonumber\\&&
       + {H}^{'V}(1,\chi_{46},\chi_{14},\chi_{16},\chi_{1}) \, \Big( 1/4\,(\chi_{6} + \chi_{4})\,\chi_{1} \Big)
       + {H}^{'V}(1,\chi_\eta,\chi_{14},\chi_{14},\chi_{1}) \, \Big( 1/24\,\chi_{1}\,\chi_\eta 
\nonumber\\&&
       + 1/6\,\chi_{1}\,\chi_\eta\,{R^z_{\eta 61}}  + 1/6\,\chi_{1}\,\chi_\eta\,{R^z_{\eta 61}} ^2 \Big)
+ {H}^{'V}(1,\chi_\eta,\chi_{16},\chi_{16},\chi_{1}) \, \Big( 1/12\,\chi_{1}\,\chi_\eta - 1/6\,\chi_{1}\,\chi_\eta\,{R^z_{\eta 61}} 
\nonumber\\&&
 + 1/12\,\chi_{1}\,\chi_\eta\,{R^z_{\eta 61}} ^2 \Big)
       + {H}^{'V}(2,\chi_{1},\chi_{1},\chi_{1},\chi_{1}) \, \Big( 2/9\,\chi_{1}^2\,{R^z_{146\eta}} \,{R^c_{146\eta}}  \Big)
\nonumber\\&&
       + {H}^{'V}(2,\chi_{1},\chi_{1},\chi_\eta,\chi_{1}) \, \Big(  - 4/9\,\chi_{1}^2\,{R^z_{146\eta}} \,{R^z_{\eta 61}} ^2 \Big)
       + {H}^{'V}(2,\chi_{1},\chi_{14},\chi_{14},\chi_{1}) \, \Big(  - 5/12\,\chi_{1}^2\,{R^z_{146\eta}}  \Big)
\nonumber\\&&
       + {H}^{'V}(2,\chi_{1},\chi_{16},\chi_{16},\chi_{1}) \, \Big(  - 5/24\,\chi_{1}^2\,{R^z_{146\eta}}  \Big)
       + {H}^{'V}(5,\chi_{1},\chi_{1},\chi_{1},\chi_{1}) \, \Big( 1/9\,\chi_{1}^2\,{R^z_{146\eta}} ^2 \Big)
\nonumber\\&&
       + {H}_1^{'V}(1,\chi_{1},\chi_{14},\chi_{14},\chi_{1}) \, \Big( 2/9\,\chi_{1}\,{R^z_{146\eta}}  + 2/3\,\chi_{1}^2\,{R^c_{146\eta}}  + 2/9\,(\chi_{4} - 4\,\chi_{1})\,\chi_{1}\,{R^z_{16\eta}}  \Big)
\nonumber\\&&
       + {H}_1^{'V}(1,\chi_{1},\chi_{16},\chi_{16},\chi_{1}) \, \Big( 1/9\,\chi_{1}\,{R^z_{146\eta}}  + 1/3\,\chi_{1}^2\,{R^c_{146\eta}}  + 1/9\,(\chi_{6} - 4\,\chi_{1})\,\chi_{1}\,{R^z_{14\eta}}  \Big)
\nonumber\\&&
       + {H}_1^{'V}(1,\chi_{14},\chi_{14},\chi_\eta,\chi_{1}) \, \Big( 2/9\,(\chi_\eta - \chi_{4})\,\chi_{1} + 8/9\,(\chi_\eta + 2\,\chi_{1})\,\chi_{1}\,{R^z_{\eta 61}} ^2 + 4/9\,(2\,\chi_\eta - \chi_{4} 
\nonumber\\&&
         + 2\,\chi_{1})\,\chi_{1}\,{R^z_{\eta 61}}  \Big)
       + {H}_1^{'V}(1,\chi_{16},\chi_{16},\chi_\eta,\chi_{1}) \, \Big( 4/9\,(\chi_\eta - \chi_{6})\,\chi_{1} + 4/9\,(\chi_\eta + 2\,\chi_{1})\,\chi_{1}\,{R^z_{\eta 61}} ^2 
\nonumber\\&&
         - 4/9\,(2\,\chi_\eta - \chi_{6} + 2\, \chi_{1})\,\chi_{1}\,{R^z_{\eta 61}}  \Big)
       + {H}_1^{'V}(2,\chi_{1},\chi_{14},\chi_{14},\chi_{1}) \, \Big( 2/3\,\chi_{1}^2\,{R^z_{146\eta}}  \Big)
\nonumber\\&&
       + {H}_1^{'V}(2,\chi_{1},\chi_{16},\chi_{16},\chi_{1}) \, \Big( 1/3\,\chi_{1}^2\,{R^z_{146\eta}}  \Big)
       + {H}_{21}^{'V}(1,\chi_{1},\chi_{14},\chi_{14},\chi_{1}) \, \Big( 1/2\,\chi_{1}^2\,{R^z_{16\eta}}  - 1/4\,\chi_{1}^2\,{R^c_{146\eta}}  \Big)
\nonumber\\&&
       + {H}_{21}^{'V}(1,\chi_{1},\chi_{16},\chi_{16},\chi_{1}) \, \Big( 1/4\,\chi_{1}^2\,{R^z_{14\eta}}  - 1/8\,\chi_{1}^2\,{R^c_{146\eta}}  \Big)
       + {H}_{21}^{'V}(1,\chi_{4},\chi_{14},\chi_{14},\chi_{1}) \, \Big( 9/8\,\chi_{1}^2 \Big)
\nonumber\\&&
       + {H}_{21}^{'V}(1,\chi_{46},\chi_{14},\chi_{16},\chi_{1}) \, \Big( 3/2\,\chi_{1}^2 \Big)
       + {H}_{21}^{'V}(1,\chi_\eta,\chi_{14},\chi_{14},\chi_{1}) \, \Big( 1/8\,\chi_{1}^2 + 1/2\,\chi_{1}^2\,{R^z_{\eta 61}} 
\nonumber\\&&
        + 1/2\,\chi_{1}^2\,{R^z_{\eta 61}} ^2 \Big) + {H}_{21}^{'V}(1,\chi_\eta,\chi_{16},\chi_{16},\chi_{1}) \, \Big( 1/4\,\chi_{1}^2 - 1/2\,\chi_{1}^2\,{R^z_{\eta 61}}  + 1/4\,\chi_{1}^2\,{R^z_{\eta 61}} ^2 \Big)
\nonumber\\&&
       + {H}_{21}^{'V}(2,\chi_{1},\chi_{14},\chi_{14},\chi_{1}) \, \Big(  - 1/4\,\chi_{1}^2\,{R^z_{146\eta}}  \Big)
       + {H}_{21}^{'V}(2,\chi_{1},\chi_{16},\chi_{16},\chi_{1}) \, \Big(  - 1/8\,\chi_{1}^2\,{R^z_{146\eta}}  \Big)
\nonumber\\&&
       + {H}_{27}^{'V}(1,\chi_{1},\chi_{14},\chi_{14},\chi_{1}) \, \Big(  - 1/2\,\chi_{1}\,{R^z_{16\eta}}  + 1/4\,\chi_{1}\,{R^c_{146\eta}}  \Big)
\nonumber\\&&
       + {H}_{27}^{'V}(1,\chi_{1},\chi_{16},\chi_{16},\chi_{1}) \, \Big(  - 1/4\,\chi_{1}\,{R^z_{14\eta}}  + 1/8\,\chi_{1}\,{R^c_{146\eta}}  \Big)
       + {H}_{27}^{'V}(1,\chi_{4},\chi_{14},\chi_{14},\chi_{1}) \, \Big(  - 9/8\,\chi_{1} \Big)
\nonumber\\&&
       + {H}_{27}^{'V}(1,\chi_{46},\chi_{14},\chi_{16},\chi_{1}) \, \Big(  - 3/2\,\chi_{1} \Big)
       + {H}_{27}^{'V}(1,\chi_\eta,\chi_{14},\chi_{14},\chi_{1}) \, \Big(  - 1/8\,\chi_{1} - 1/2\,\chi_{1}\,{R^z_{\eta 61}} 
\nonumber\\&&
        - 1/2\,\chi_{1}\,{R^z_{\eta 61}} ^2 \Big) + {H}_{27}^{'V}(1,\chi_\eta,\chi_{16},\chi_{16},\chi_{1}) \, \Big(  - 1/4\,\chi_{1} + 1/2\,\chi_{1}\,{R^z_{\eta 61}}  - 1/4\,\chi_{1}\,{R^z_{\eta 61}} ^2 \Big)
\nonumber\\&&
       + {H}_{27}^{'V}(2,\chi_{1},\chi_{14},\chi_{14},\chi_{1}) \, \Big( 1/4\,\chi_{1}\,{R^z_{146\eta}}  \Big)
       + {H}_{27}^{'V}(2,\chi_{1},\chi_{16},\chi_{16},\chi_{1}) \, \Big( 1/8\,\chi_{1}\,{R^z_{146\eta}}  \Big)
\end{eqnarray}

\end{document}